\documentclass[acmsmall]{acmart}
\usepackage{soul}
\usepackage{enumitem}
\usepackage{mathtools}
\usepackage{empheq}
\usepackage[normalem]{ulem}
\usepackage{tikz}
\usetikzlibrary{arrows.meta,positioning,calc,fit,backgrounds}

\setcopyright{cc}
\setcctype{by}
\copyrightyear{2026}
\acmYear{2026}
\acmJournal{PACMNET}
\acmVolume{4}
\acmNumber{CoNEXT2}
\acmArticle{24}
\acmMonth{6}
\acmDOI{10.1145/3808672}

\received{December 2025}
\received[accepted]{April 2026}

\newcommand{\hlmone}[1]{#1}
\newcommand{\hlmtwo}[1]{#1}
\newcommand{\hlmthree}[1]{#1}

\graphicspath{{figures/}}
\usepackage{subfigure}
\usepackage{xspace}
\newcommand{\ours}{\textsc{SWOT}\xspace}
\newcommand{\newapproach}{intra-collective reconfiguration\xspace}
\newcommand{\newapproachst}{intra-collective\xspace}
\newcommand{\strawman}{Strawman-ICR\xspace}
\newcommand{\us}{\textmu{}s\xspace}

\authorsaddresses{} 
\begin{document}
\title[SWOT: Enabling Comm-Reconf Overlap for CC in Optical Networks]{SWOT: Enabling Communication-Reconfiguration Overlap for Collective Communication in Optical Networks}
\newcommand{\socs}{School of Computer Science and Technology, USTC}
\newcommand{\siar}{Suzhou Institute for Advanced Research, USTC}

\author{Changbo Wu}
\email{wuchangbo@mail.ustc.edu.cn}
\orcid{0009-0004-9285-0594}
\affiliation{
  \institution{School of Computer Science and Technology, University of Science and Technology of China (USTC)}
  \city{Hefei}
  \state{Anhui}
  \country{China}
}
\affiliation{
  \institution{Shanghai Innovation Institute}
  \city{Shanghai}
  \country{China}
}
\affiliation{
  \institution{\siar}
  \city{Suzhou}
  \state{Jiangsu}
  \country{China}
}

\author{Zhuolong Yu}
\authornote{Corresponding author. Email address: zhuolongyu@ustc.edu.cn.}
\email{zhuolongyu@ustc.edu.cn}
\orcid{0000-0002-8846-5229}

\author{Gongming Zhao}
\email{gmzhao@ustc.edu.cn}
\orcid{0000-0003-1311-8908}

\author{Hongli Xu}
\email{xuhongli@ustc.edu.cn}
\orcid{0000-0003-3831-4577}
\affiliation{
  \institution{\socs}
  \city{Hefei}
  \state{Anhui}
  \country{China}
}
\affiliation{
  \institution{\siar}
  \city{Suzhou}
  \state{Jiangsu}
  \country{China}
}



\begin{abstract}
Collective communication (CC) is critical for scaling distributed machine learning (DML). The predictable traffic patterns of DML present a great opportunity for applying optical network technologies.
Optical networks with reconfigurable topologies promise high bandwidth and low latency for collective communications.
However, existing approaches face inherent limitations: static topologies are inefficient for dynamic communication patterns within CC algorithms, while frequent topology reconfiguration matching every step of the algorithm incurs significant overhead.

In this paper, we propose \ours, a demand-aware optical network framework that employs ``\newapproach'' to dynamically align network resources with CC traffic patterns.  \ours hides reconfiguration latency by overlapping it with data transmission through three key techniques: \textit{Heterogeneous Message Splitting}, \textit{Asynchronous Overlapping}, and \textit{Topology Bypassing}.
Extensive simulations demonstrate that \ours reduces communication completion time up to 89.7\% across diverse CC algorithms compared to static baselines, demonstrating strong robustness to varying optical resources and reconfiguration delay.

\end{abstract}



\maketitle
\section{Introduction}

The scaling laws\cite{kaplanScalingLawsNeural2020} dictate that the AI model size and training data size are critical factors determining the model capability. To achieve high model performance and capability, distributed machine learning (DML) has emerged as an essential strategy.
Efficient DML relies on distributed computing clusters with high bandwidth, low end-to-end latency, and high scalability\cite{jiangMegaScaleScalingLarge2024a,wangDomainSpecificNetworkTransport2024}. 

In recent decades, significant investments have driven the development of optical network technologies, with optical circuit switches (OCSs) now being widely deployed in modern data centers\cite{ballaniSiriusFlatDatacenter2020,jouppiTPUV4Optically2023,zuResiliencyScaleManaging2024a,poutievskiJupiterEvolvingTransforming2022,liuLightwaveFabricsAtScale2023}.
In DML training, the communication pattern is dominated by predictable, high-throughput collective communication (CC) operations, which makes it particularly well suited for leveraging the reconfigurability and high-capacity switching of optical networks\cite{ghobadiEmergingOpticalInterconnects2022,khaniSiPMLHighbandwidthOptical2021,jouppiTPUV4Optically2023,wangTopoOptCooptimizingNetwork2023}. However, existing approaches are constrained either spatially, by the limited optical degree of static topologies, or temporally, by the non-negligible latency of optical reconfiguration.

Current optical interconnect-based CC schemes (\textit{e.g.}, TopoOpt~\cite{wangTopoOptCooptimizingNetwork2023}) predominantly adopt a ``one-shot'' reconfiguration strategy.
By establishing a fixed topology that persists throughout the collective operation, these approaches effectively avoid the overhead introduced by reconfiguring the optical network.
However, this static paradigm faces a fundamental scalability barrier: modern DML workloads increasingly rely on more efficient but complex CC implementation (\textit{e.g.}, Pairwise All-to-All Algorithm, Rabenseifner's algorithm for AllReduce) with highly dynamic, time-varying connectivity demands.
Embedding the union of these diverse demands into a single static optical topology often exceeds the physical optical degree of the network.
Consequently, systems are compelled to rely on suboptimal static-friendly algorithms or indirect multi-hop routing, causing non-negligible performance degradation and compromising training efficiency.

To effectively support these dynamic communication patterns, {an alternative direction} is to enable ``\newapproach'' \textemdash~dynamically adapt network topology to match the specific demand of each algorithmic step.
However, the naive implementation of this approach (which we refer to as \textit{Strawman}) faces a harsh reality: optical switching is not instant.
Standard OCS devices incur a non-negligible latency $T_{\text{recfg}}$ to complete topology reconfiguration once.
If the network strictly synchronizes reconfiguration with every communication step, the system enters a ``stop-and-wait'' cycle.
For multi-step algorithms, the accumulated reconfiguration overhead often outweighs the advanced CC algorithm benefits, rendering this approach impractical.



We argue that the key to unlocking the potential of optical networks lies in enabling ``\newapproach'' without paying the full reconfiguration latency cost.
To this end, we propose \ours, a demand-aware optical network framework that makes ``\newapproach'' practical.
\ours incorporates three key mechanisms:
(1) \textbf{Heterogeneous Message Splitting:} It strategically adjusts message splits to create temporal windows for overlap and bypassing;
(2) \textbf{Asynchronous Overlapping:} It removes strict synchronization across parallel optical planes, enabling some planes/OCSs to reconfigure while others continue transmitting;
(3) \textbf{Topology Bypassing:} It selectively bypasses intermediate reconfiguration stages for certain planes, directly eliminating switching overheads.
Our contributions are summarized as follows:
\begin{itemize}
    \item We identify fundamental limitations in the current optical CC design space, where static strategies are constrained by limited optical degrees and naive dynamic strategies incur prohibitive reconfiguration latency.
    \item We propose \ours, a scheduling framework that leverages message splitting, overlap, and bypassing techniques to mask reconfiguration latency.
    \item We perform a comprehensive evaluation covering efficiency and scalability across CC primitives, and hardware sensitivity. Results show \ours reduces communication time up to 89.7\% across diverse primitives and exhibits robustness against varying optical degrees ($k$) and reconfiguration latencies ($T_{\text{recfg}}$).
\end{itemize}

\begin{figure}[t]
    \centering
    \vspace{-0.2in}
    \begin{minipage}[t]{0.35\textwidth}
        \centering
        \includegraphics[width=0.9\linewidth]{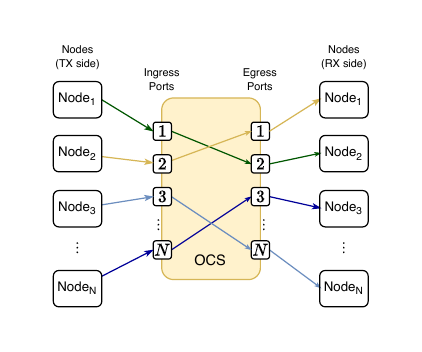}
        \vspace{-0.1in}
        \captionof{figure}{Example of OCS interconnect. Left and right sides denote the same nodes (TX and RX paths, respectively).}
        \Description{A schematic of an optical circuit switch interconnect. The same set of nodes appears on the left and right; left-side ports represent transmit paths and right-side ports represent receive paths. Lines through the switch show a one-to-one circuit mapping between transmit and receive ports.}
        \label{subfig:interconnect}
    \end{minipage}
    \hfill
    \begin{minipage}[t]{0.62\textwidth}
        \centering
        \includegraphics[width=\linewidth]{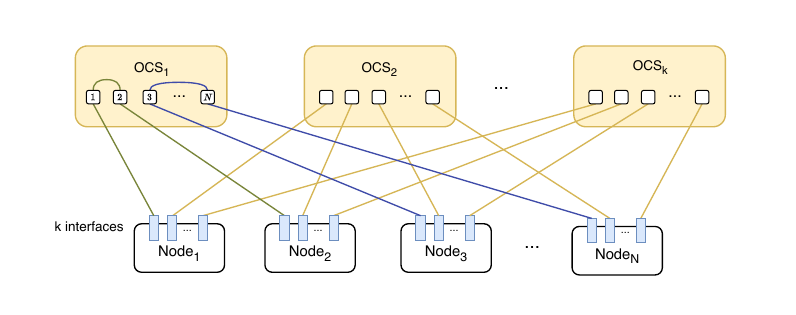}
        \vspace{-0.15in}
        \captionof{figure}{Example of optical network topology, where OCS$_1$ is reconfigured to match Fig.~\ref{subfig:interconnect}.}
        \label{subfig:topo}
    \end{minipage}
    \vspace{-0.1in}
\end{figure}

\section{Background \& Motivation}
\label{sec:background_motivation}
\subsection{Background}
\subsubsection{Optical Network and Optical Circuit Switch}

Modern optical networks deliver \textit{tera-scale bandwidth}  and \textit{deterministic $\mu s$-level latency} through direct photonic propagation, bypassing electronic packet processing bottlenecks.
Among optical switching technologies, Optical Circuit Switching (OCS) achieves this by creating an optical path, or ``circuit'', between the source and destination.
As shown in Fig~\ref{subfig:interconnect}, an $N \times N$ OCS establishes a one-to-one mapping between its ingress and egress ports.
This functionality can be realized using different physical technologies, such as MEMS mirrors, liquid crystal or thermo-optic switches. In a MEMS-based OCS, a representative example, the connectivity is physically realized by an $N{\times}N$ mirror array that physically steers optical beams to link the designated ingress and egress ports.
Notably, this intrinsic \textbf{bijective mapping} is a fundamental property of OCS regardless of hardware, and can be uniformly formalized as a permutation matrix $\mathbf{P} \in \{0,1\}^{N \times N}$ where $p_{ij}=1$ iff input $i$ connects to output $j$.

A defining feature of OCS is its \textbf{reconfigurability} -- the ability to dynamically change the permutation $\mathbf{P}$ to match evolving traffic patterns.
However, this flexibility comes with a switching overhead, denoted as \textbf{reconfiguration latency} ($T_{\text{recfg}}$).
This latency is determined by the underlying physical technology.
Recent hardware advances have expanded OCS reconfiguration latency to span $10$ ns \cite{ballaniSiriusFlatDatacenter2020}, $10$ $\mu$s \cite{melletteRotorNetScalableLowcomplexity2017}, and $10$ ms.
However, optical technologies face a fundamental tradeoff among three key parameters: port-count, reconfiguration latency, and insertion loss \cite{ghobadiEmergingOpticalInterconnects2022}.
Specifically, OCS devices achieving ns-to-\us-scale reconfiguration times typically suffer from either severely limited port counts or high insertion loss, making \us-to-ms-scale OCS devices more practical for large-scale GPU cluster interconnects.
This makes the overhead of reconfiguration non-negligible during collective operations in optical-interconnected clusters.

Fig.~\ref{subfig:topo} illustrates the physical view of the direct-connect optical topology adopted in this work.
The cluster consists of $N$ computing nodes interconnected through $k$ parallel OCS devices.
Each node possesses $k$ optical interfaces, where the $j$-th interface of every node is physically connected to a distinct port on the $j$-th OCS ($OCS_j$).
This structure effectively creates $k$ parallel optical planes, where $OCS_j$ is responsible for establishing connections between the $j$-th interfaces of any pair of nodes.
By configuring the internal port mappings of these $k$ OCSs, specific logical topologies (e.g., rings, tori, or random graphs) can be constructed over the physical infrastructure.

The optical degree $k$ is a critical parameter defining the network's parallelism.
It can be scaled by equipping nodes with additional physical network adapters or, more commonly, by leveraging the \textit{Port Splitting} (or \textit{Breakout}) capabilities inherent in modern high-performance NICs.
Port splitting allows a single high-bandwidth physical port to be logically reconfigured into multiple independent, lower-bandwidth interfaces to meet diverse networking demands.
For instance, the NVIDIA C8180 ConnectX-8 SuperNIC~\cite{PortSplittingConfigurations}, featuring a single-port OSFP module capable of 800 Gb/s, can be reconfigured from its default configuration to support two 400 GbE ports ($k$=2) or up to eight 100 GbE ports ($k$=8).
Similar capabilities are widely supported by other adapters, such as the Intel Ethernet 800 Series~\cite{IntelEthernet800}, Broadcom NetXtreme~\cite{BroadcomEthernetNetwork}, and Marvell FastLinQ 45000~\cite{EthernetAdaptersControllers}.


\subsubsection{Collective Communication} \label{subsub: cc}
Modern DML workloads increasingly adopt hybrid parallelism strategies to handle the heavy computational load, combining {data parallelism}, {tensor parallelism}, {expert parallelism}{, context parallelism}, etc\cite{yeDeepLearningWorkload2024}.  
 These parallelization schemes rely on collective communication (CC) primitives (\textit{e.g.}, AllReduce, AllGather, All-to-All) as the coordination backbone for synchronizing gradients, parameters, and intermediate tensors across distributed accelerators.

It is crucial to distinguish between \textbf{CC primitives} and \textbf{CC algorithms}.
CC primitives (\textit{e.g.}, AllReduce, AllGather, All-to-All) define the high-level communication semantics---specifying what data needs to be aggregated or exchanged.
Each primitive can be implemented by multiple \textit{CC algorithms} (\textit{e.g.}, Ring, Tree, Recursive-Doubling, Bruck), each optimized for different message sizes, node counts ($p$), and network topologies.

The performance of these algorithms is typically modeled using the $\alpha-\beta$ cost model~\cite{hockney1994communication}, where $\alpha$ represents the fixed per-message latency (including software overhead and propagation delay) and $\beta$ represents the per-byte transmission time (inverse of bandwidth).
Despite its simplicity, this model remains widely accepted in recent state-of-the-art literature~\cite{sandersTwotreeAlgorithmsFull2009, xutingliuRethinkingMachineLearning2024, wangBlinkFastGeneric2020, caiSynthesizingOptimalCollective2021, shahTACCLGuidingCollective2023, sensiSwingShortcuttingRings2024, caoSyCCLExploitingSymmetry2025, zhaoEfficientDirectConnectTopologies2025} and has been empirically verified via linear regression on physical testbeds~\cite{zhaoEfficientDirectConnectTopologies2025}, confirming its fidelity for modeling modern high-speed interconnects.

Different algorithms dominate different regimes based on their $\alpha$ and $\beta$ complexities.
For example, Ring-based algorithms are bandwidth-optimal (minimized $\beta$ term) but suffer from high latency scaling linearly with $p$, making them suitable for large messages.
Conversely, algorithms like Bruck or Recursive-Doubling optimize latency ($\mathcal{O}(\log p)$ steps) but may induce bandwidth contention, making them ideal for small messages.
Production libraries (\textit{e.g.}, NCCL, MPI) dynamically select the \textit{Best-Known Algorithm} based on runtime conditions to minimize the estimated completion time.

\begin{figure}[t]
  \vspace{0.1in}
  \centering
  \includegraphics[width=0.71\columnwidth]{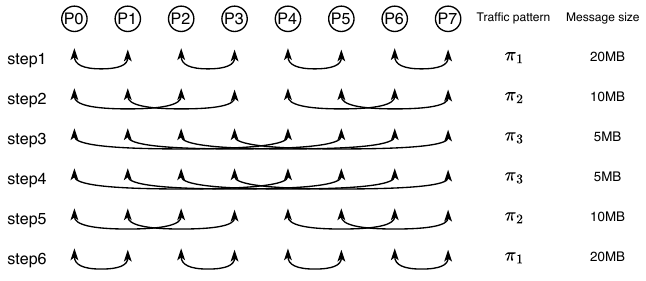}
  \caption{For an 8-node AllReduce with a 40~MB payload, Rabenseifner's algorithm proceeds in six communication steps. Each node splits the data into eight 5~MB segments. Steps~1--3 implement a reduce-scatter using recursive halving with three bijective pairings, after which node~$P_i$ owns the reduced $i$-th segment. Steps~4--6 perform an allgather using recursive doubling to disseminate these segments to all nodes. The \emph{Message size} column on the right shows the amount of data exchanged at each stage of every step.}
  \Description{A six-step diagram for Rabenseifner's AllReduce on eight nodes. The first three steps show recursive-halving reduce-scatter pairings, and the last three steps show recursive-doubling allgather pairings. The rightmost column lists the message volume exchanged at each step, decreasing during reduce-scatter and increasing during allgather.}
  \vspace{-0.25in}
  \label{fig: allreduce-hd}
\end{figure}


A key characteristic of CC lies in its \textbf{multi-step execution}: each collective algorithm decomposes into sequential communication steps where every node participates in exclusively pairwise data transfers at each step.
This avoids simultaneous many-to-one or one-to-many traffic patterns that create congestion hotspots.
The multi-step communication patterns can be formalized as sequences of \textbf{bijective pairings}. For $N=8$ nodes labeled $\{0,\dots,7\}$, let $\pi_t: [N] \to [N]$ denote the pairing function at step $t$, where each node $i$ communicates with a peer $\pi_t(i)$.
Crucially, these steps are strictly sequential: step $t+1$ relies on the data reduced or collected in step $t$, creating a global synchronization barrier between steps.

\textbf{AllReduce (Ring):} At each step, node $i$ sends its data to $\pi_t(i) = (i + 1) \bmod 8$, propagating partial reductions through adjacent nodes in a circular pipeline. 

\textbf{AllReduce (Rabenseifner's Algorithm\cite{rabenseifnerOptimizationCollectiveReduction2004}):} Fig.~\ref{fig: allreduce-hd} shows the pattern of $\log_2 N$-step ($N=8$) Reduce-Scatter phase.
\begin{itemize}
\item \textit{Step 1:} $\pi_1(i) = i \oplus 1$
  (2-node microgroups: $\{0{\leftrightarrow}1,\ 2{\leftrightarrow}3,\ 4{\leftrightarrow}5,\ 6{\leftrightarrow}7\}$)
\item \textit{Step 2:} $\pi_2(i) = i \oplus 2$
  (4-node subgroups: $\{0{\leftrightarrow}2,\ 1{\leftrightarrow}3,\ 4{\leftrightarrow}6,\ 5{\leftrightarrow}7\}$)
\item \textit{Step 3:} $\pi_3(i) = i \oplus 4$
  (Cross-group: $\{0{\leftrightarrow}4,\ 1{\leftrightarrow}5,\ 2{\leftrightarrow}6,\ 3{\leftrightarrow}7\}$)
\end{itemize}
The Allgather phase reverses this pattern in the next $\log_2 N$ steps. 

\textbf{All-to-All (Pairwise Exchange):}
This algorithm employs $N-1$ steps to progressively disseminate data blocks:
At each step $t \in [1,2,...,N-1]$, node $i$ sends the data block to $\pi_t(i)=(i + t) \bmod 8$, accumulating one new block per step.

\subsubsection{Mapping Collective Algorithms to Optical Substrates} \label{subsub: relationship}
\begin{figure}[t]
  \centering
  \includegraphics[width=0.56\columnwidth]{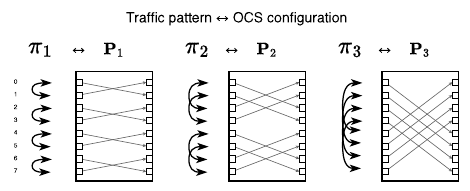}
  \vspace{-0.1in}
  \caption{OCS configuration corresponding to traffic pattern in 8-node Rabenseifner's AllReduce (see Fig~\ref{fig: allreduce-hd})}
  \Description{A mapping diagram that translates the three bijective communication pairings in the eight-node Rabenseifner AllReduce into corresponding optical circuit switch configurations. Each configuration is represented as a permutation connecting node interfaces through an OCS plane.}
  \vspace{-0.14in}
  \label{fig: mapping}
\end{figure}

The multi-step nature of CC algorithms (described in $\S$\ref{subsub: cc}) introduces a \textbf{time-varying connectivity demand}.
Formally, a collective algorithm can be viewed as a sequence of bijective pairings $\Pi = \{\pi_1, \pi_2, \dots, \pi_{|\Pi|}\}$, where $\pi_t$ represents the specific logical topology required at algorithmic step $t$.
To support this demand, the optical network provides a sequence of physical OCS configurations $\mathcal{C} = \{\mathbf{P}_1, \mathbf{P}_2, \dots, \mathbf{P}_{|\mathcal{C}|}\}$, where each $\mathbf{P}_j$ represents a set of established circuit permutations.
For instance, as illustrated in Fig.~\ref{fig: mapping}, each bijective pairing stage ($\pi_1$, $\pi_2$, $\pi_3$) requires a corresponding OCS bijective mapping ($\mathbf{P}_1$, $\mathbf{P}_2$, $\mathbf{P}_3$).

The fundamental challenge lies in embedding the logical sequence $\Pi$ into the physical sequence $\mathcal{C}$ under the constraints of the optical hardware.
A valid mapping ensures that for any algorithmic step $t$, the required connectivity $\pi_t$ is supported by the active physical configuration.
This mapping can be achieved through two distinct dimensions:
\begin{itemize}[leftmargin=*]
    \item \textbf{Spatial Reservation:} Provisioning a physical topology that simultaneously contains links for multiple (or all) logical steps. This relies on the spatial parallelism (optical degree $k$) to accommodate the union of demands.
    \item \textbf{Temporal Reconfiguration:} Dynamically switching the physical topology over time to match the changing logical demands step-by-step. This relies on the temporal agility ($T_{\text{recfg}}$) of the OCS.
\end{itemize}

\subsection{Current Solutions and Limitations} \label{sec:constraints}
Based on the mapping dimensions discussed in $\S$\ref{subsub: relationship}, the current design space for optical collective communication is polarized, with practical solutions clustering at the static end.

\textbf{1. The Status Quo: Static Pre-configuration (``One-shot''):}
Approaches like TopoOpt~\cite{wangTopoOptCooptimizingNetwork2023} prioritize spatial reservation, provisioning a single, static topology for the entire collective operation, and even the whole training iteration (\textit{i.e.}, $|\mathcal{C}|=k$, where $k$ is the optical degree, typically $k \in \{2, 4, 8\}$).
To ensure connectivity throughout the collective operation without reconfiguration, this static configuration $\bigcup_{i=1}^{k} \mathbf{P}_{i}$ must structurally embed the \textit{union} of all logical pairings ($\bigcup_{t=1}^{|\Pi|} \pi_t$).
While effective for sparse patterns like Ring-AllReduce, this approach fails for complex algorithms.
For instance, the Pairwise-Exchange All-to-All algorithm implies a union graph equivalent to a fully connected clique ($K_N$).
Supporting such a dense graph on nodes with limited constant-level $k$ ($k \ll N$) is physically impossible.
Consequently, ``One-shot'' schemes are compelled to either discard communication-efficient algorithms that exhibit dynamic patterns in favor of suboptimal static-friendly variants, or rely on indirect multi-hop routing to satisfy connectivity.
Both approaches inevitably lead to significant performance degradation compared to direct transmission.

\textbf{2. The Hypothetical Alternative: Naive Intra-Collective Reconfiguration.}
To overcome the limitations of static topologies, one might consider exploiting the temporal dimension by strictly switching the configuration to match each algorithmic step (\textit{i.e.}, $|\mathcal{C}| = |\Pi|$).
We refer to this hypothetical baseline as the naive \newapproach (\strawman).
While this maximizes spatial efficiency by dedicating all optical links to the current demand, it suffers from prohibitive temporal overhead.
Since standard OCS devices incur a reconfiguration latency $T_{\text{recfg}}$ (\textit{e.g.}, $200 \mu s$), pausing communication at every step introduces a cumulative ``stop-and-wait'' penalty of $|\Pi| \times T_{\text{recfg}}$.
As shown in our later analysis ($\S$\ref{sec:moti_example}), this latency penalty often outweighs the bandwidth benefits, rendering the strategy impractical for sequences with many steps.

Existing solutions face inherent physical constraints of OCSs: static topologies are limited by optical degree (spatial constraint), while naive dynamic approaches are bottlenecked by reconfiguration latency (temporal constraint). SWOT aims to bridge this gap by exploring the design space in between---dynamically scheduling partial reconfigurations and overlapping them with data transmission---thereby making intra-collective reconfiguration practical.

\section{\ours Design}
\subsection{Design Intuition} \label{sec:moti_example}

\begin{figure*}[tbp]
    \centering
  \includegraphics[width=\textwidth]{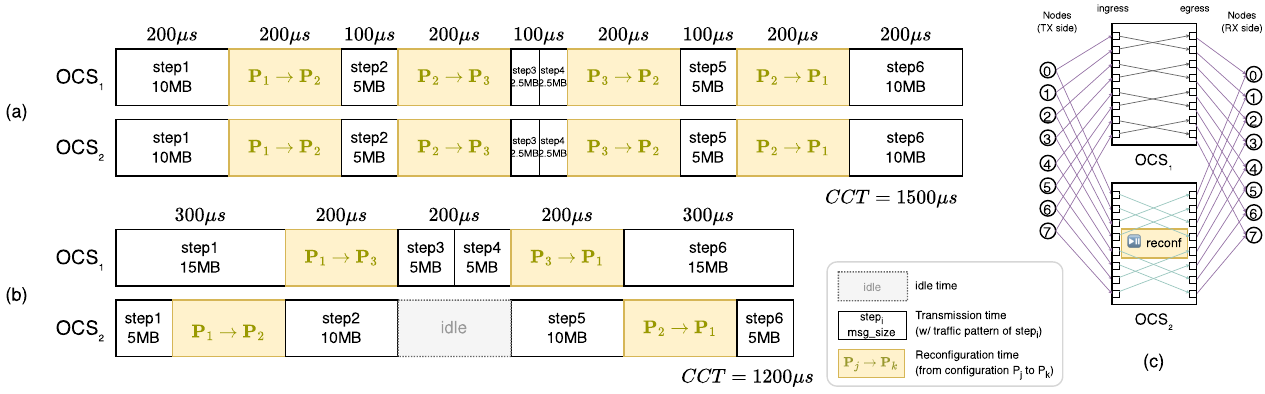}
 \vspace{-0.3in}
 \caption{Design intuition behind overlapping communication with reconfiguration.
  Communication-reconfiguration timeline for 8-node AllReduce in optical network:
    (a) naive \newapproach incurs cumulative $800$ \us switching overhead,
    (b) \ours's overlap-optimized approach reduces CCT by 20\% through partial circuit updates during transmissions (Illustrated in (c)).
  Configuration details ($400$Gbps links, $200$ \us reconfiguration delay), traffic patterns and required OCS configurations shown in Fig.~\ref{fig: allreduce-hd} and Fig.~\ref{fig: mapping}.}
  \Description{A three-part timeline diagram for an eight-node AllReduce over two optical circuit switch planes. Part (a) shows a naive intra-collective reconfiguration schedule in which both OCS planes transmit and reconfigure in lockstep, accumulating four visible reconfiguration gaps. Part (b) shows the SWOT schedule, where traffic is split unevenly so one plane can reconfigure while the other continues transmitting, reducing the total completion time. Part (c) highlights the partial circuit-update intuition that enables reconfiguration work to overlap with data transmission.}
  \vspace{-0.25in}
  \label{fig: moti_example}
\end{figure*}


To overcome the limitations highlighted in \S\ref{sec:constraints}, \ours employs a demand-aware scheduling paradigm.
We revisit the 8-node Rabenseifner's AllReduce example from \S\ref{subsub: cc}, with \ours and the naive \newapproach approaches.
Fig.~\ref{fig: moti_example} depicts the corresponding timelines under $k=2$ OCS planes.
Analysis of the traffic patterns in Fig.~\ref{fig: mapping} reveals that the collective operation requires a specific sequence of topological configurations: $\mathbf{P}_1 \to \mathbf{P}_2 \to \mathbf{P}_3 \to \mathbf{P}_3 \to \mathbf{P}_2 \to \mathbf{P}_1$, which directly correspond to the Reduce-Scatter and AllGather phases.

\textbf{The Inefficiency of Lockstep Scheduling (Fig.~\ref{fig: moti_example}(a)).}
Fig.~\ref{fig: moti_example}(a) depicts the timeline of the naive \newapproach (\strawman).
In this paradigm, the two OCS planes operate in \textit{lockstep}: they serve the same communication step simultaneously and reconfigure synchronously.
For instance, at Step 1, the 20MB traffic is evenly split between $OCS_1$ and $OCS_2$ (10MB each), taking 200 $\mu s$ to transmit.
Crucially, before Step 2 can begin, \textit{both} OCSs must stop and reconfigure from $\mathbf{P}_1$ to $\mathbf{P}_2$, incurring a global penalty of $T_{\text{recfg}} = 200 \mu s$.
This ``stop-and-wait'' cycle repeats for every topological transition.
Consequently, the total Communication Completion Time (CCT) is dominated by the accumulated reconfiguration overhead:
$CCT = \sum t_{trans} + 4 \times T_{\text{recfg}} = 700 \mu s + 800 \mu s = 1500 \mu s$.
Here, reconfiguration accounts for 53.3\% of the total time, severely limiting system efficiency.
\textit{(Note: In this example, we omit the end-to-end base latency $T_{\text{lat}}$ for simplicity, as it is negligible compared to both reconfiguration overhead $T_{\text{recfg}}$ and transmission time.)
}


\begin{table}[t]
\caption{Summary of Key Notations}
\vspace{-0.16in}
\label{tab:notations}
\begin{tabular}{p{1.1cm}p{0.55cm}p{11.3cm}}
\toprule
\textbf{Symbol} & \textbf{Type} & \textbf{Description} \\
\midrule
\multicolumn{3}{l}{\textit{Decision Variables}} \\
\hlmthree{$d_{i,j}$} & \hlmthree{$\mathbb{R}^+$} & \hlmthree{Message/Data volume assigned to OCS $j$ at step $i$} \\
\hlmthree{$r_{i,j}$} & \hlmthree{$\{0,1\}$} & \hlmthree{1 if OCS $j$ needs reconfiguration at step $i$} \\
\midrule
\multicolumn{3}{l}{\textit{Intermediate Variables}} \\
$u_{i,j}$ & $\{0,1\}$ & 1 if OCS $j$ is used at step $i$, 0 otherwise \\
$s_{i,j}$ & $\{0,1\}$ & 1 if OCS $j$'s current config matches step $i$\\
\hlmthree{$t_{\text{start}_{i,j}}$} & \hlmthree{$\mathbb{R}^+$} & \hlmthree{Transmission start time on OCS $j$ at step $i$} \\
\hlmthree{$t_{\text{end}_{i,j}}$} & \hlmthree{$\mathbb{R}^+$} & \hlmthree{Transmission end time on OCS $j$ at step $i$} \\
\hlmthree{$t_{\text{recfg\_s}_{i,j}}$} & \hlmthree{$\mathbb{R}^+$} & \hlmthree{Reconfig start time for OCS $j$ at step $i$} \\
\hlmthree{$t_{\text{recfg\_e}_{i,j}}$} & \hlmthree{$\mathbb{R}^+$} & \hlmthree{Reconfig end time for OCS $j$ at step $i$} \\
$t_{\text{prev\_e}_{i,j}}$ & $\mathbb{R}^+$ & last completion time of previous activities (trans / reconf) in OCS $j$ before step $i$ \\
$t_{\text{step\_e}_i}$ & $\mathbb{R}^+$ & Completion time of communication step $i$ \\
$\text{last\_cfg}_{i,j}$ & $\mathbb{N}$ & \hlmthree{Configuration ID held by OCS $j$ before step $i$} \\
\midrule
\multicolumn{3}{l}{\textit{Parameters}} \\
$m_i$ & $\mathbb{R}^+$ & Total data volume required at step $i$ \\
$B$ & $\mathbb{R}^+$ & OCS port bandwidth (Gbps) \\
$T_{\text{lat}}$ & $\mathbb{R}^+$ & End-to-end base latency (propagation + processing) \\
$T_{\text{recfg}}$ & $\mathbb{R}^+$ & \hlmone{Conservative path-readiness latency for OCS reconfiguration} \\
$M$ & $\mathbb{R}^+$ & Large constant value for big-M method \cite{cococcioniBigMMethodNumerical2021} \\
$\text{cfg}_i$ & $\mathbb{N}$ & \hlmthree{Configuration ID required by step $i$} \\
\bottomrule
\end{tabular}
\vspace{-0.25in}
\end{table}

\textbf{Optimization via Asynchrony and Bypassing (Fig.~\ref{fig: moti_example}(b)).}
The root cause of the inefficiency above is the rigid synchronization.
\ours breaks this rigid synchronization, achieving performance gains through three synergistic mechanisms:
\begin{enumerate}
\item \textbf{Heterogeneous Message Splitting}: \ours strategically splits the message and distributes the workload unevenly across OCS planes. For example, in step 1, it assigns a larger chunk (15MB) to $OCS_1$ and a smaller chunk (5MB) to $OCS_2$.

    \item \textbf{Asynchronous Overlapping:} The uneven split creates a time window. $OCS_2$ finishes its small chunk early (at $t=100 \mu s$) and immediately begins reconfiguring. By the time $OCS_1$ finishes Step 1 (at $t=300 \mu s$), $OCS_2$ has already completed the switch to $\mathbf{P}_2$ and is ready to transmit. This effectively ``hides'' the latency of $OCS_2$ behind the transmission of $OCS_1$.
    \hlmone{At runtime, the corresponding operations are released by completion and path-ready events rather than fixed wall-clock triggers.}

   \item \textbf{Topology Bypassing:}
    Beyond overlapping, \ours identifies opportunities to \textit{skip} unnecessary intermediate configurations.
    In the naive approach, every OCS must strictly follow the sequence $\mathbf{P}_1 \to \mathbf{P}_2 \to \mathbf{P}_3 \dots$.
    However, \ours recognizes that $OCS_1$ does not need to handle Step 2 ($\mathbf{P}_2$) if $OCS_2$ can absorb that demand.
    Consequently, $OCS_1$ transitions directly from $\mathbf{P}_1 \to \mathbf{P}_3$, completely bypassing $\mathbf{P}_2$.
    Similarly, $OCS_2$ handles Steps 2 and 5 (both $\mathbf{P}_2$) while skipping $\mathbf{P}_3$.
    This optimization reduces the total number of reconfiguration operations per OCS, directly saving time overhead.
\end{enumerate}

Through this ``relay-style'' execution--where planes not only overlap tasks but also intelligently skip unnecessary reconfiguration--\ours reduces the CCT to 1200 $\mu s$, achieving a \textbf{20\% speedup} without adding physical resources.
The following sections describe how we generalize this intuition into a formal scheduling framework.

\subsection{\ours Scheduler: Formulation \& Constraints} \label{sec:schedule}

While the intuition in \S\ref{sec:moti_example} demonstrates the potential of \ours, applying it to complex algorithms and large clusters requires solving a high-dimensional combinatorial problem. The scheduler must co-optimize traffic distribution and topology transitions while strictly adhering to the physical and logical constraints of the optical cluster.

\subsubsection{Scheduling Properties}
To ensure a valid and executable schedule, our formulation must satisfy three fundamental properties:
\begin{itemize}
    \item \phantomsection\label{prop:p1}\textit{(P1) Transmission-Reconfiguration Precedence:} Data transmission on a specific OCS plane can only commence after the necessary optical circuit is fully established.
    \item \phantomsection\label{prop:p2}\textit{(P2) Serial Resource Usage:} A single OCS device cannot perform multiple activities (e.g., transmitting and reconfiguring) simultaneously.
    \item \phantomsection\label{prop:p3}\textit{(P3) Logical Step Barrier:} \hlmone{To preserve collective semantics, step $i$ data transfers wait for step $i-1$ data completion across participating planes; the barrier does not forbid earlier path preparation on free planes.}
\end{itemize}

\subsubsection{MILP Formulation}
We formulate this problem as a Mixed Integer Linear Programming (MILP) model. The complete notation is provided in Table~\ref{tab:notations}.
We introduce two sets of \textbf{decision variables} that serve as the ``control knobs'' for our design mechanisms:
\begin{itemize}
    \item \textbf{Message Split ($d_{i,j}$):} The data volume assigned to OCS $j$ at step $i$. This variable enables \textbf{Workload Distortion} by allowing $d_{i,j}$ to deviate from the uniform split ($m_i/k$).
    \item \textbf{Reconfiguration Flag ($r_{i,j}$):} A binary variable indicating if OCS $j$ reconfigures before step $i$. This variable enables \textbf{Configuration Bypassing} ($r_{i,j}=0$) to skip unnecessary latency.
\end{itemize}
\hlmthree{The scheduler's core decisions are traffic allocation $d_{i,j}$ and reconfiguration choice $r_{i,j}$. The timestamp variables in Table~\ref{tab:notations} certify a feasible ordering and CCT under properties \hyperref[prop:p1]{(P1)}--\hyperref[prop:p3]{(P3)}; the runtime consumes the induced dependencies as event guards, not clock-triggered commands.}


\textbf{Optimization Model.}
The objective is to minimize the Communication Completion Time (CCT): $\min \ \text{CCT} = \max_{i} t_{\text{step\_e}_i}$.
The constraints are formulated to enforce properties \hyperref[prop:p1]{(P1)}--\hyperref[prop:p3]{(P3)} while enabling asynchronous execution:

\begin{empheq}[left = \empheqlbrace]{align}
\sum\nolimits_{j=1}^{k} d_{i,j} \times u_{i,j} &= m_i \quad \forall i \label{eq:traffic} \\
t_{\text{end}_{i,j}} - t_{\text{start}_{i,j}} &= \frac{d_{i,j}}{B} + T_{\text{lat}} \cdot u_{i,j} \quad \forall i,j \\
t_{\text{recfg\_e}_{i,j}} - t_{\text{recfg\_s}_{i,j}} &= r_{i,j} \cdot T_{\text{recfg}} \quad \forall i,j \label{eq:reconf_dur} \\
r_{i,j} &\geq u_{i,j} - s_{i,j} \quad \forall i,j \label{eq:reconf_logic_1}\\
| \text{cfg}_i - \text{last\_cfg}_{i,j} | &\leq M \cdot (1 - s_{i,j}) \quad \forall i,j \label{eq:reconf_logic_2}\\
t_{\text{prev\_e}_{1,j}} &= 0 \\
t_{\text{prev\_e}_{i,j}}
&\ge
\max\left\{
t_{\text{prev\_e}_{i-1,j}},
t_{\text{end}_{i-1,j}} \cdot u_{i-1,j},
t_{\text{recfg\_e}_{i-1,j}} \cdot r_{i-1,j}
\right\}
\quad \forall i>1,j
\label{eq:timeline} \\
t_{\text{recfg\_s}_{i,j}} &\geq t_{\text{prev\_e}_{i,j}} \label{eq:async_start} \\
\hlmthree{t_{\text{start}_{i,j}}} &\hlmthree{\geq t_{\text{recfg\_e}_{i,j}} - M(1-u_{i,j}) \quad \forall i,j} \label{eq:p1_precedence} \\
\hlmthree{t_{\text{step\_e}_i}} &\hlmthree{\geq t_{\text{end}_{i,j}} - M(1-u_{i,j}) \quad \forall i,j} \label{eq:step_end} \\
t_{\text{start}_{i,j}} &\geq t_{\text{step\_e}_{i-1}} \quad \forall i>1,j \label{eq:p3_sync}
\end{empheq}

\subsubsection{Constraint Analysis}
We analyze how these constraints enforce rigorous correctness while unlocking the proposed optimizations:

\textbf{1. Flexible Assignment via Traffic Conservation (Eq.~\ref{eq:traffic}).}
Eq.~(\ref{eq:traffic}) ensures that the total message $m_i$ is distributed among active OCSs. Unlike static schemes that enforce equality, this constraint gives the solver the freedom to adjust $d_{i,j}$ arbitrarily. This mathematical flexibility is the foundation of \textbf{Workload Distortion}, allowing the scheduler to sacrifice the transmission time of one plane to buy reconfiguration time for another.

\textbf{2. Decoupled Timelines via Serial Resource Usage (Eq.~\ref{eq:reconf_dur}--\ref{eq:p1_precedence}).}
These constraints manage the internal timeline of each OCS $j$ to satisfy properties \textit{(P1)} and \textit{(P2)}:
\begin{itemize}
    \item \textbf{Enabling Bypassing:} Eq.~(\ref{eq:reconf_logic_1}) and (\ref{eq:reconf_logic_2}) govern the reconfiguration logic. \hlmthree{Since configurations are encoded as IDs, Eq.~(\ref{eq:reconf_logic_2}) simply tests whether OCS $j$ already holds the configuration required by step $i$.} They allow $r_{i,j}=0$ (and thus zero latency penalty in Eq.~\ref{eq:reconf_dur}) if the current physical configuration matches the requirement ($s_{i,j}=1$), effectively mathematically realizing \textbf{Configuration Bypassing}.
    \item \textbf{Enabling Asynchronous Overlap:} Eq.~(\ref{eq:timeline}) and (\ref{eq:async_start}) define the earliest start time for a new operation based \textit{only} on the previous activity of the \textit{same} OCS. Crucially, Eq.~(\ref{eq:async_start}) permits $OCS_j$ to start reconfiguring immediately after its previous task finishes ($t_{\text{prev\_e}_{i,j}}$), regardless of the state of other OCS planes. This local dependency (as opposed to global synchronization) is what formally enables \textbf{Asynchronous Overlap}.
    \item \textbf{Ensuring Integrity:} Eq.~(\ref{eq:p1_precedence}) enforces \textit{(P1)}, guaranteeing that transmission ($t_{\text{start}}$) never occurs before the circuit is ready ($t_{\text{recfg\_e}}$) \hlmthree{for active OCS planes ($u_{i,j}=1$), while the big-M term relaxes the constraint for inactive planes}.
\end{itemize}

\textbf{3. \hlmone{Logical Step Barrier} (Eq.~\ref{eq:step_end}--\ref{eq:p3_sync}).}
While operations within a step are asynchronous, Eq.~(\ref{eq:step_end}) and (\ref{eq:p3_sync}) enforce \textit{(P3)} as a logical step barrier.
\hlmone{They preserve the collective algorithm structure by forcing step $i$ data transfers to wait for the completion of step $i-1$ across active planes, while still allowing path preparation to proceed earlier on free planes.}
\hlmthree{Eq.~(\ref{eq:step_end}) only binds active planes to the step completion time; inactive planes are relaxed by the big-M term.}
This represents the boundary condition for our optimization: we maximize overlap \textit{inside} the steps to minimize the waiting time at these barriers.

\subsubsection{Solving Feasibility}
\label{sec:feasibility}
While MILP is computationally intensive, it does not constitute a bottleneck for \ours in production DML environments. This is supported by three key factors:
\begin{enumerate}
    \item \textbf{Offline Amortization:} DML workloads are characterized by highly deterministic and repetitive communication patterns~\cite{jiangMegaScaleScalingLarge2024a}. Consequently, \ours functions as a one-time \textit{offline} planner. The cost of schedule generation is incurred only during initialization and is amortized over millions of iterations in a training job (often spanning weeks).
    \item \textbf{Relaxed Optimality:} Our objective is not to find the global optimum, but one that significantly outperforms the static and naive baselines.  MILP solvers typically find high-quality feasible solutions early in the search process, spending most time closing the ``optimality gap''. In practice, a near-optimal solution that overlaps most of the reconfiguration latency is often sufficient to capture the majority of the end-to-end benefit.
    \item \textbf{Empirical Efficiency:} In our extensive evaluation (see \S\ref{sec:evaluate}), we utilized an off-the-shelf open-source solver (CBC via Pulp\cite{mitchellPulpLinearProgramming2011}) with a strict time limit of 120 seconds.
Empirical results demonstrate that the solver consistently converges to solutions yielding significant performance gains within this short window, even for large-scale clusters (\textit{e.g.}, $p=1024$).  This confirms that standard solvers are sufficiently efficient for practical deployment without requiring specialized heuristic algorithms.
\end{enumerate}

\subsection{\ours Framework \& Execution Model}
\label{subsec: framework}

\begin{figure}[htbp]
  \vspace{0.15in}
  \centering
  \includegraphics[width=0.8 \textwidth]{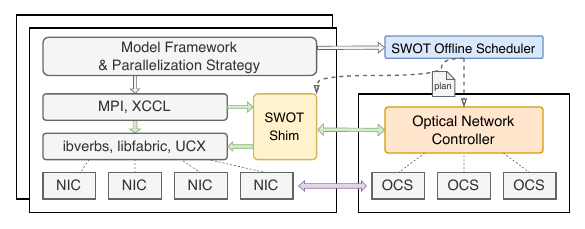}
  \vspace{-0.1in}
  \caption{\textbf{System Integration of \ours.} The framework consists of an Offline Scheduler for plan generation and a runtime Shim layer. The schedule is installed into both the Shim and optical controller; at runtime, the Shim coordinates transfers with controller path-readiness events.}
  \Description{A system architecture diagram for SWOT. An offline scheduler takes predictable collective communication patterns and produces a schedule package. The schedule is installed into host-side Shim components and into an optical controller. At runtime, the Shim coordinates data transfers with the communication library or NIC workers, while the optical controller manages OCS path preparation and returns path-readiness events used to release guarded transfers.}
  \vspace{-0.1in}
  \label{fig: shim_arch}
\end{figure}

\hlmone{Designed to bridge offline collective scheduling with dynamic optical hardware, \ours operates in two phases: offline schedule generation and runtime event-driven coordination.
The offline phase identifies traffic-splitting and reconfiguration decisions, while the runtime phase lowers the resulting schedule into operations released by readiness and completion events.}

\hlmone{\textbf{Offline Scheduling.}
As shown in Figure~\ref{fig: shim_arch}, \ours first runs an offline scheduler that extracts the predictable collective pattern and solves the MILP in \S\ref{sec:schedule}.
The output is a global schedule package that records, for each step $i$ and OCS plane $j$, the traffic assignment ($d_{i,j}$), reconfiguration decision ($r_{i,j}$), and dependency order.
During initialization, this package is distributed into runtime views for the host-side Shims and the optical controller.}

\hlmone{\textbf{Runtime Execution.}
At runtime, the Shim releases operations through event guards rather than clock-triggered commands.
Path preparation may start when its OCS plane is free, while a transfer is issued only after its data dependency is ready, the required optical path is confirmed, and previous conflicting operations on that plane have completed.
This event-driven execution implements properties \hyperref[prop:p1]{(P1)}--\hyperref[prop:p3]{(P3)} without relying on fine-grained clock synchronization~\cite{gengExploitingNaturalNetwork2018,namyarFireflyScalableUltraAccurate2025}: delayed events may reduce realized overlap, but dependent operations simply wait.}

\hlmone{\textbf{Compatibility and Boundary.}
The Shim is a mediation layer alongside standard communication libraries rather than a replacement for their low-level transports.
It lowers collective steps into P2P transfers on per-plane communicators and assumes predictable collective schedules, reliable completion events, a stable out-of-band control channel, and a conservative path-ready event.
Accordingly, $T_{\text{recfg}}$ denotes path-readiness latency, including switch settling or platform-specific link-readiness delays when relevant.
Our evaluation validates the schedule-level benefit under this execution contract, but does not measure a production Shim/controller implementation or its full software and control-plane overhead.}

\hlmone{Appendix~\ref{app:execution-model-details} illustrates this event-driven execution model, including how it handles path readiness, completion-driven progress, and runtime variability.}

\section{Evaluation} \label{sec:evaluate}
We comprehensively evaluate \ours to demonstrate its efficiency and scalability across different CC algorithms, with direct comparisons for each primitive.
We also investigate its performance under varying network settings, examining the impact of optical resources and reconfiguration speeds on the scheduling strategy.

\subsection{Experimental Setup}
We evaluate \ours through rigorous numerical simulations.
Similar to~\cite{zhaoEfficientDirectConnectTopologies2025}, the communication time is calculated using the $\alpha-\beta$ model, which accounts for both the fixed latency ($\alpha$) and bandwidth-dependent transmission time ($\beta$).
This approach ensures that our evaluation captures realistic transmission behaviors, including serialization overheads and propagation delays, rather than relying solely on ideal bandwidth models.

\textbf{Network Topology.} We simulate a representative direct-connect optical topology (Fig.~\ref{subfig:topo}) consisting of $p$ computing nodes interconnected by $k$ OCSs.
Each node is equipped with $k$ interfaces, each connected to a distinct OCS, so the evaluated substrate exposes $k$ parallel and independently reconfigurable optical planes.
\hlmtwo{Accordingly, the experiments isolate the schedule-level benefits of \ours on this representative topology; extensions to routed multi-dimensional optical fabrics and dynamic sparse workloads are discussed in \S\ref{sec:discussion}.}
Following common commercial designs~\cite{xueOpticalSwitchingData2023}, we set the OCS reconfiguration delay $T_{\text{recfg}}$ to $200~\mu$s.
The impact of reconfiguration latency is further evaluated in \S~\ref{subsec:impact_reconf}.


\textbf{Network Parameters.} To reflect modern hardware capability, we set the total aggregate bandwidth per computing node at 800 Gbps. Consequently, the per-link bandwidth $B$ is determined by the degree of parallelism, calculated as $B = 800 / k $ Gbps.
$k$ stands for the optical degree, which leverages the breakout capabilities inherent in modern high-performance NICs.
To account for realistic physical layer overheads, we set the end-to-end base latency $T_{\text{lat}}$ to $20 \mu s$, which captures the typical propagation delay and hardware serialization time in data center optical interconnects. 

\textbf{Solver Configuration.}
For the optimization component of \ours, we employ the Pulp optimizer to solve the MILP formulation.
To ensure practical feasibility and fair comparison across all experiments, all schedules are generated with a strict solver time limit of 120 seconds, unless explicitly stated otherwise.
This constraint emulates the requirement for rapid schedule generation in production environments.

\textbf{Algorithm Selection.} For the algorithm-level evaluation ($\S$~\ref{subsec:perf_over_alg} and $\S$~\ref{subsec:scale_over_alg}), we select four representative CC algorithms: (1) Rabenseifner's ReduceScatter; (2) Rabenseifner's AllReduce; (3) Pairwise All-to-All; and (4) Bruck's All-to-All~\cite{bruckEfficientAlgorithmsAlltoall1997}.
For the primitive-level evaluation ($\S$~\ref{subsec:perf_over_primitive}), we adopt a ``best-of-breed'' strategy: we automatically select the collective algorithm that minimizes the estimated CCT for each scheme, ensuring a fair comparison against the best-case static configurations.

\textbf{Baseline Methods.} \ours is compared against three scheduling paradigms:
    (1) \textbf{One-shot}: \hlmthree{Static pre-installation of optical circuits before a collective starts; it realizes only the communication configurations that fit within the available OCS planes and does not reconfigure during execution};
    (2) \textbf{\strawman}: Naive \newapproach without overlap optimization~\cite{athapathuReconfigurabilityCollectiveCommunication2025};
    (3) \textbf{Ideal}: Communication without network constraints at maximum aggregate NIC bandwidth.

\subsection{Performance Analysis}
In this section, we conduct a multi-faceted evaluation to quantify the benefits of \ours. We structure our analysis across three key dimensions:
\begin{enumerate}[leftmargin=*]
    \item \textbf{Algorithmic Efficiency:} We examine how \ours accelerates specific representative algorithms across varying message sizes, highlighting its ability to reduce CCT through overlap scheduling.
    \item \textbf{Cluster Scalability:} We assess the system's behavior as the cluster size scales, demonstrating \ours's advantage in handling increasingly complex traffic patterns.
    \item \textbf{Primitive Performance Envelope:} We evaluate \ours at the primitive level (\textit{e.g.,} AllReduce, All-to-All). By selecting the best-performing algorithm for each schedule method and network configuration, we demonstrate how \ours expands the overall performance envelope for fundamental primitives, independent of the underlying algorithmic implementation.
\end{enumerate}

\subsubsection{Collective Operation Efficiency} \label{subsec:perf_over_alg}
\begin{figure*}[htbp]
    \centering
    \subfigure[ReduceScatter w/ Rabenseifner's]{
        \includegraphics[width=0.48\linewidth]{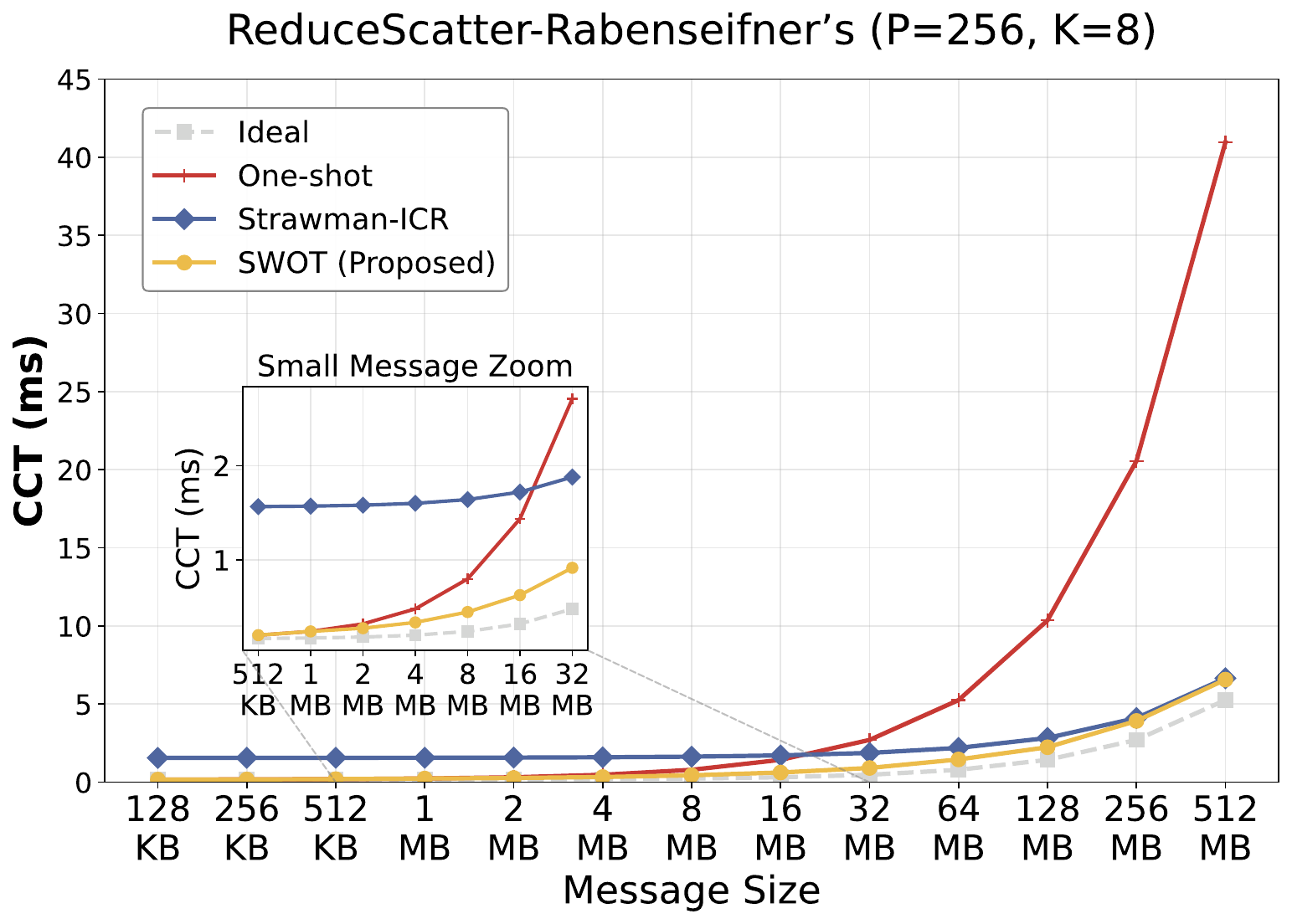}
        \label{subfig:rs_hd}
    }
    \subfigure[AllReduce w/ Rabenseifner's]{
        \includegraphics[width=0.48\linewidth]{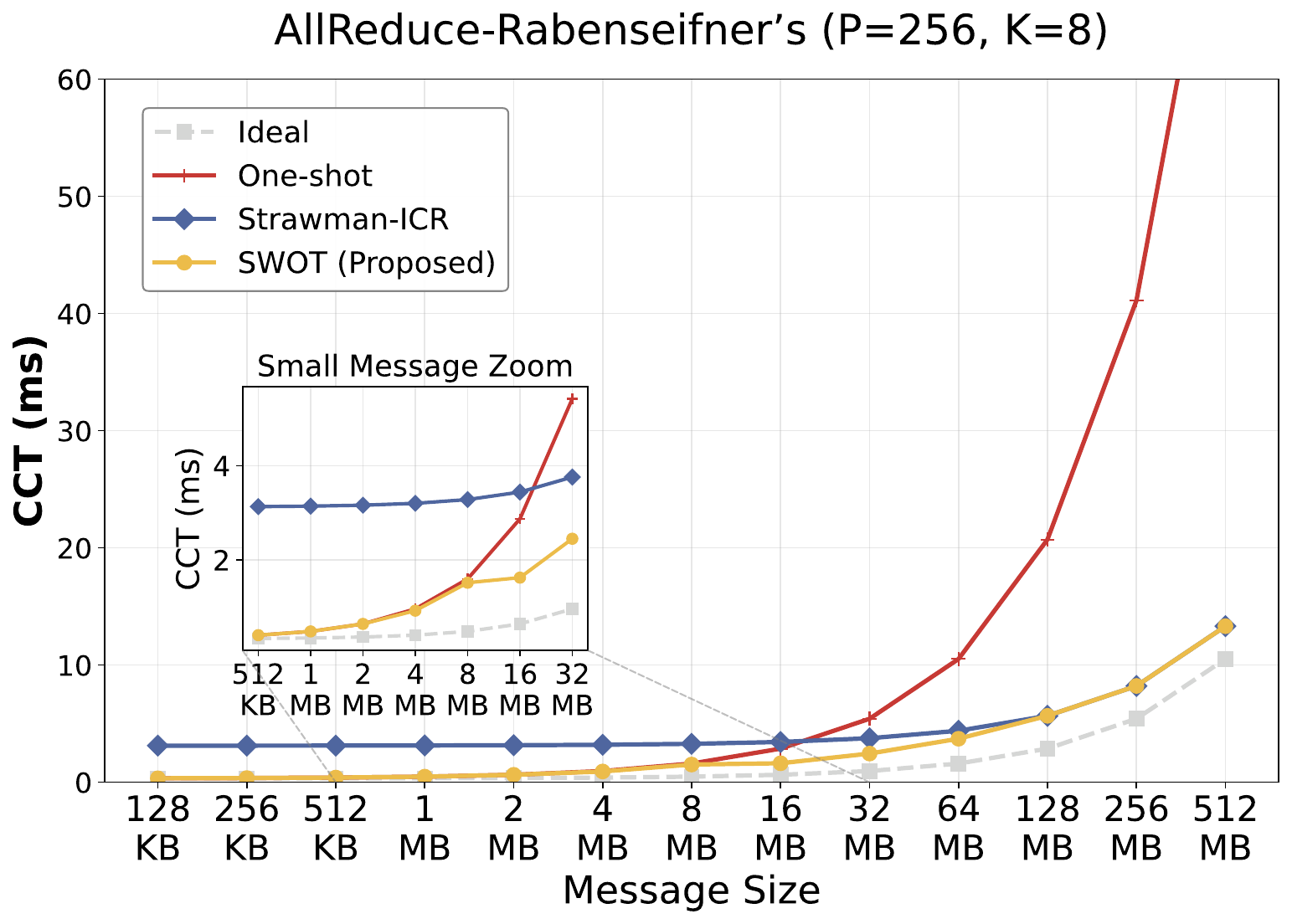}
        \label{subfig:ar_hd}
    }
    \vspace{-0.1in}
    \subfigure[All-to-All w/ Bruck]{
        \includegraphics[width=0.48\linewidth]{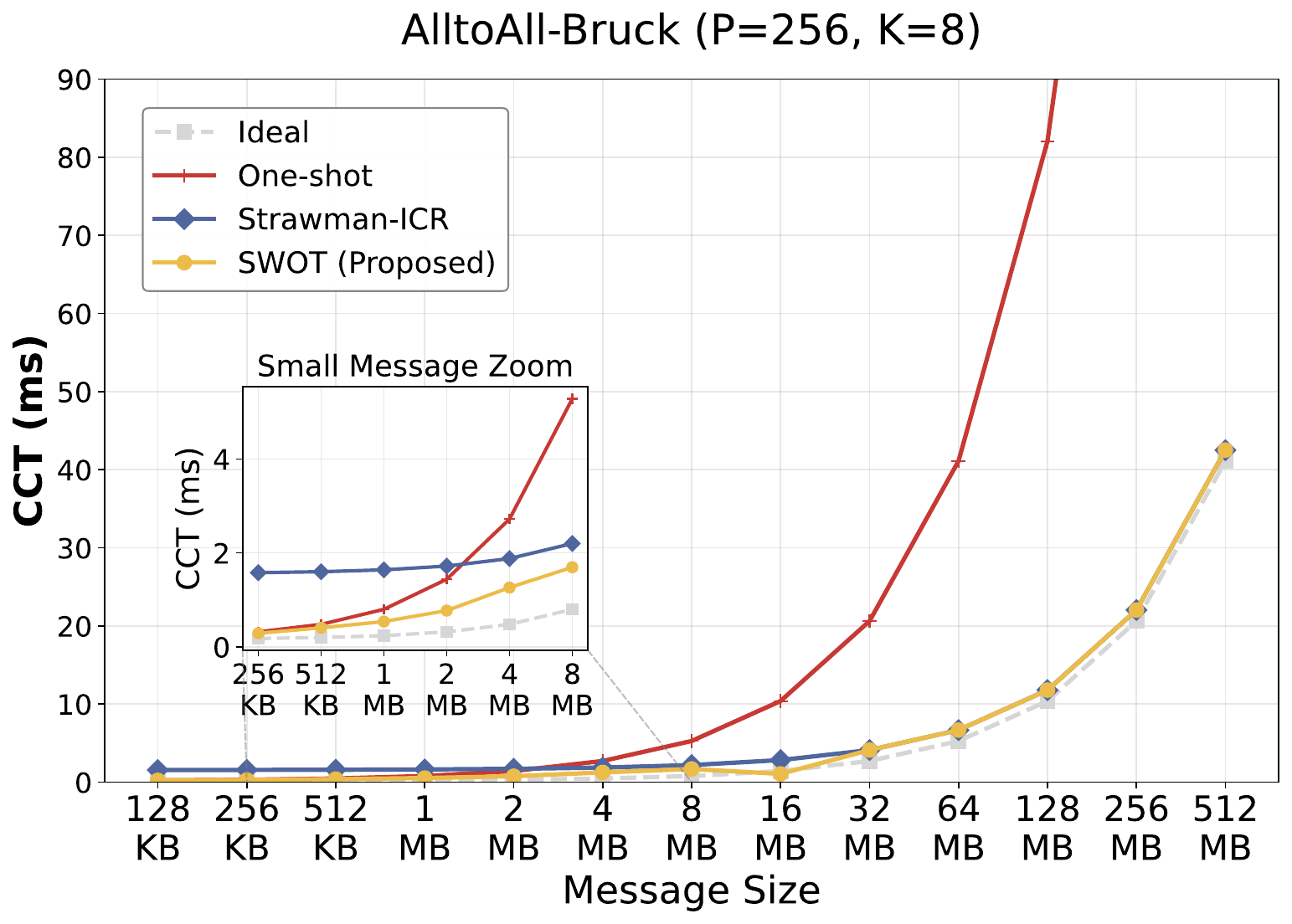}
        \label{subfig:a2a_bruck}
    }
    \subfigure[All-to-All w/ Pairwise]{
        \includegraphics[width=0.48\linewidth]{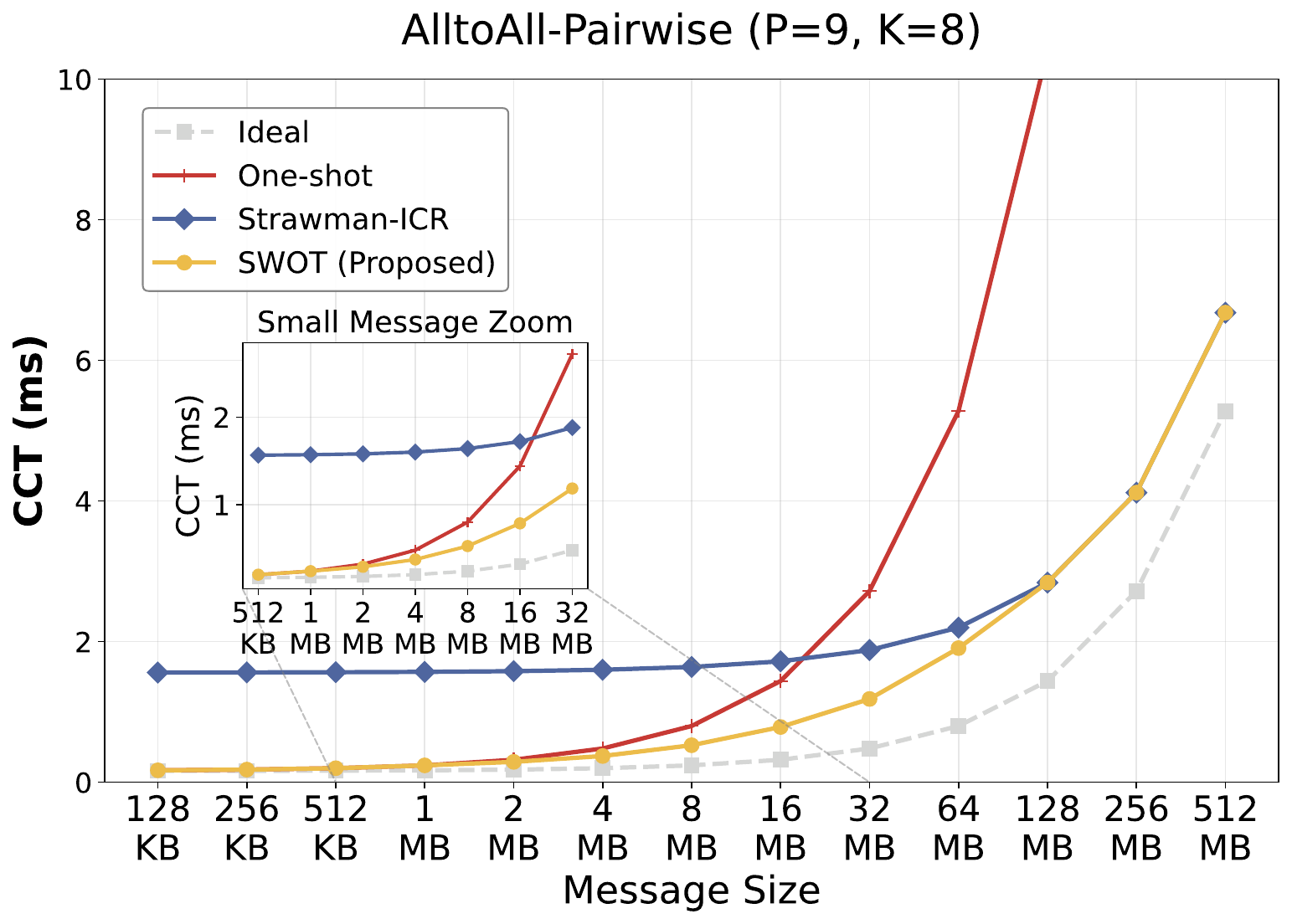}
        \label{subfig:a2a_pair}
    }
    \caption{CCT vs message size for different collective operations algorithms on a dedicated cluster of 256 nodes physically fully connected to 8 OCSs ($B = 100$ Gbps). 9-node setup for Pairwise due to one-shot scalability constraints. The small plot in the bottom left reports runtime for small message results.}
    \Description{Four performance plots compare communication completion time against message size for ReduceScatter with Rabenseifner's algorithm, AllReduce with Rabenseifner's algorithm, All-to-All with Bruck, and All-to-All with Pairwise. Each plot compares One-shot, Strawman-ICR, SWOT, and Ideal baselines. SWOT generally tracks closer to Ideal and below Strawman-ICR, with the largest advantages in regimes where reconfiguration overhead can be overlapped with transmission.}
    \label{fig: cct_msg}
\end{figure*}

Fig.~\ref{fig: cct_msg} compares \ours against existing solutions across different CC algorithms.
Compared to one-shot, \ours reduces CCT by up to 84.0\%, 83.7\%, 72.7\%, and 89.7\% for Rabenseifner's ReduceScatter, AllReduce, Pairwise All-to-All, and Bruck's All-to-All, respectively;
similarly, \ours outperforms \strawman by up to 89.1\%, 87.1\%, 89.1\%, and 85.4\%, respectively.
These results demonstrate \ours's superiority in accelerating diverse CC algorithms over optical networks.
Note that there is a gap between \ours and the ideal scenario because \ours must reserve time for optical reconfiguration, and the link bandwidth is not 100\% utilized.

Specifically, the experimental results reveal three key observations.
(1) \textbf{\ours and \strawman achieve sublinear scaling}, whereas one-shot shows linear CCT growth as message size increases.
This occurs because one-shot's static pre-allocation activates only the subset of OCSes matching the traffic pattern, wasting the bandwidth of idle links.
In contrast, dynamic reconfiguration enables higher network utilization across the communication steps.
(2) \textbf{Reconfiguration overhead is non-negligible for small messages.} For small messages ( $<1$ MB in Fig.~\ref{subfig:a2a_pair}), both \strawman and \ours exhibit comparable or higher CCT than one-shot. Reconfiguration overhead rivals transmission duration here, where naive \newapproach scheduling incurs penalties from frequent reconfigurations. \textit{\ours alleviates this via overlapped reconfiguration-communication technique.}
(3) \textbf{The performance gap between \strawman and \ours narrows with large messages} ($>64$MB in Fig~\ref{subfig:a2a_pair}),
as data transmission time becomes the dominant factor.

Furthermore, we observe that \textbf{\ours yields varying improvements depending on the collective algorithm}.
Bruck's All-to-All (Fig.~\ref{subfig:a2a_bruck}) shows relatively lower gains despite higher total data volume due to its limited number of communication phases, which restricts reconfiguration opportunities. In contrast, Pairwise All-to-All (Fig.~\ref{subfig:a2a_pair}) and Rabenseifner's AllReduce (Fig.~\ref{subfig:ar_hd}) exhibit distinct levels of improvement.
Although they share identical data volumes, their distinct configuration sequences and message distributions account for the variation in benefits.

\subsubsection{Cluster Scalability} \label{subsec:scale_over_alg}
\begin{figure*}[htbp]
  \centering
  \subfigure[AllReduce w/ Rabenseifner's]{
      \includegraphics[width=0.48\columnwidth]{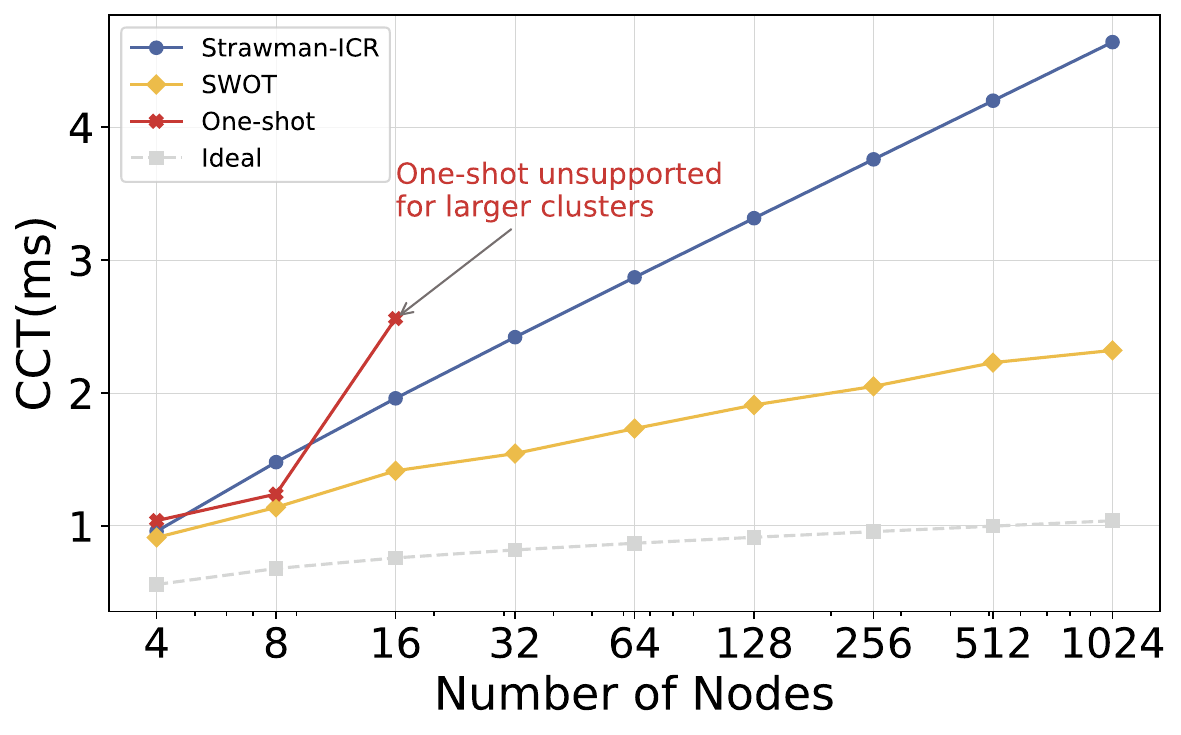}
\label{subfig:scale-cct-ar}
  }
  \subfigure[All-to-All w/ Pairwise]{
\includegraphics[width=0.48\columnwidth]{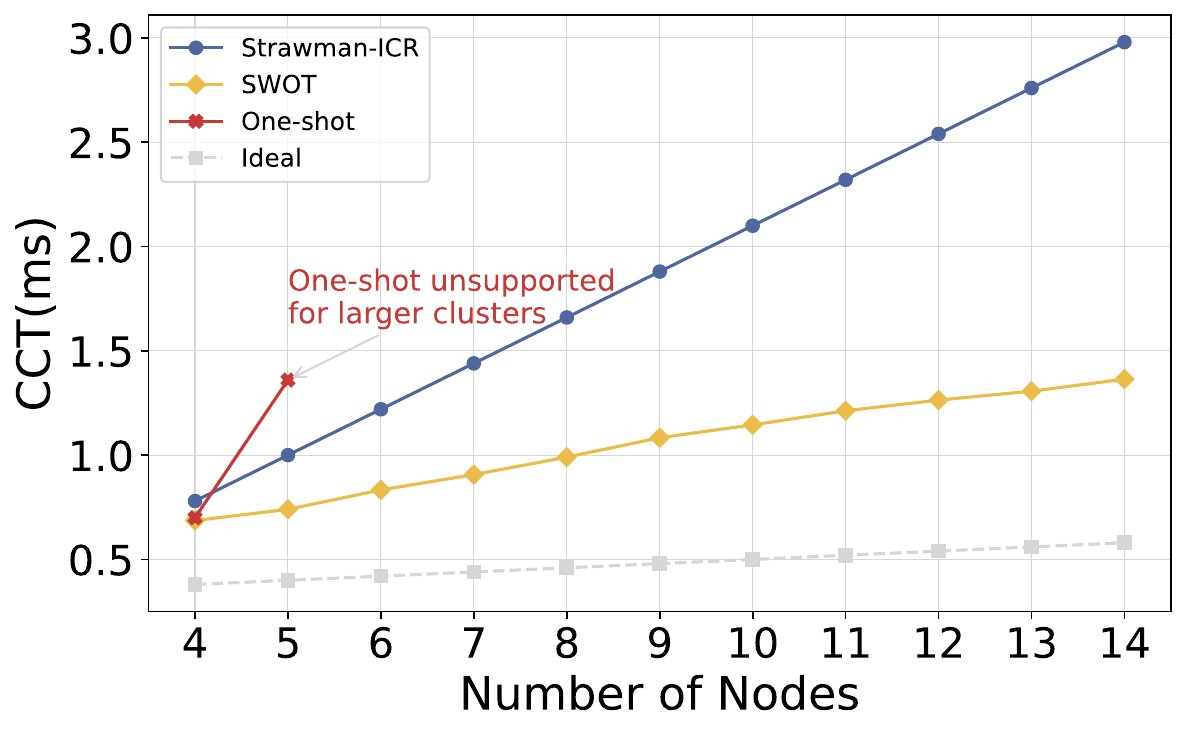}
\label{subfig:a2a_pa}
  }
  \vspace{-0.15in}
  \caption{Impact of cluster size on CCT for different CC algorithm on a dedicated cluster physically fully connected to 4 OCSs ($B = 200$ Gbps, message size = $32$ MB).}
  \Description{Two plots show communication completion time as cluster size increases for Rabenseifner AllReduce and Pairwise All-to-All with four OCS planes. One-shot is available only for smaller cluster sizes before its static configuration capacity is exceeded. Strawman-ICR and SWOT continue across larger sizes, and SWOT remains below Strawman-ICR with an increasing gap as the communication patterns become more diverse.}
  \vspace{-0.1in}
  \label{fig: scale-cct}
\end{figure*}

Fig.~\ref{fig: scale-cct} shows CCT scaling with cluster size for Rabenseifner's AllReduce and Pairwise's All-to-All operations using 4 OCSs. Key observations include:
(1) \hlmthree{One-shot supports only up to 16-node clusters for AllReduce and 5-node clusters for All-to-All because it must place all required step configurations before the collective starts.}
\hlmthree{With 4 OCSs, only four distinct circuit configurations can be kept active at once.}
\hlmthree{Pairwise All-to-All requires a different round matching for each peer exchange, while Rabenseifner's AllReduce traverses a growing sequence of partner patterns across its halving/doubling steps.}
\hlmthree{Once these required patterns exceed the available OCS planes, one-shot cannot statically realize the full algorithm.}
Both \strawman and \ours overcome this by enabling runtime reconfiguration and reusing the same OCS planes across collective steps.
(2) Larger clusters induce more diverse traffic patterns, increasing reconfiguration overhead in \strawman. \ours reduces this overhead by co-optimizing data transfer and optical reconfiguration, improving scalability.
Consequently, \ours's performance advantage over \strawman improves with cluster size: for Rabenseifner's AllReduce, CCT reduction grows from 39.6\% at 64 nodes to 46.9\% at 512 nodes; for Pairwise All-to-All, improvement rises from 26.0\% at 5 nodes to 45.5\% at 10 nodes. 
\textbf{Large-scale clusters thus derive greater benefits from \ours, highlighting its scalability for complex CC algorithms.}

\subsubsection{End-to-End Primitive Performance} \label{subsec:perf_over_primitive}
\ours is designed to be algorithm-agnostic: it does not prescribe a specific CC algorithm but rather optimizes the execution of any given algorithm on optical substrates. To validate this, we evaluate the end-to-end performance of two common primitives: AllReduce and All-to-All.

\hlmthree{Previous experiments fix representative algorithms to isolate \ours's scheduling behavior.}
\hlmthree{Here we instead evaluate primitive-level performance with a ``best-of-breed'' strategy: for each primitive/data point, we select the baseline's lowest-CCT CC algorithm, then apply \ours to measure the remaining scheduling gain.}

\begin{figure*}[htbp]
  \centering
  \subfigure[All-to-All]{
    \includegraphics[width=0.49\columnwidth]{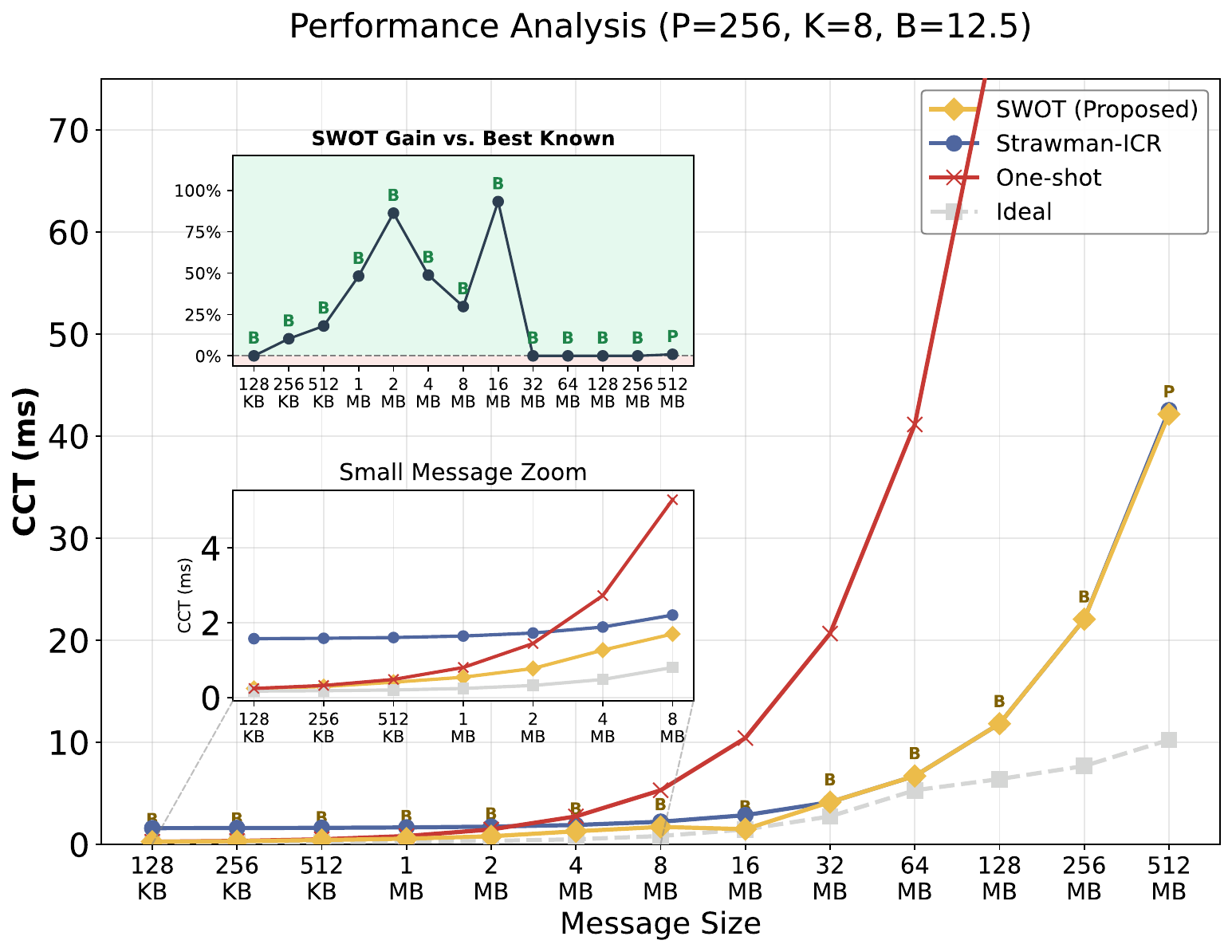}
\label{fig: primitive-a2a}
  }
  \hspace{-0.15in}
  \subfigure[AllReduce]{
      \includegraphics[width=0.49\columnwidth]{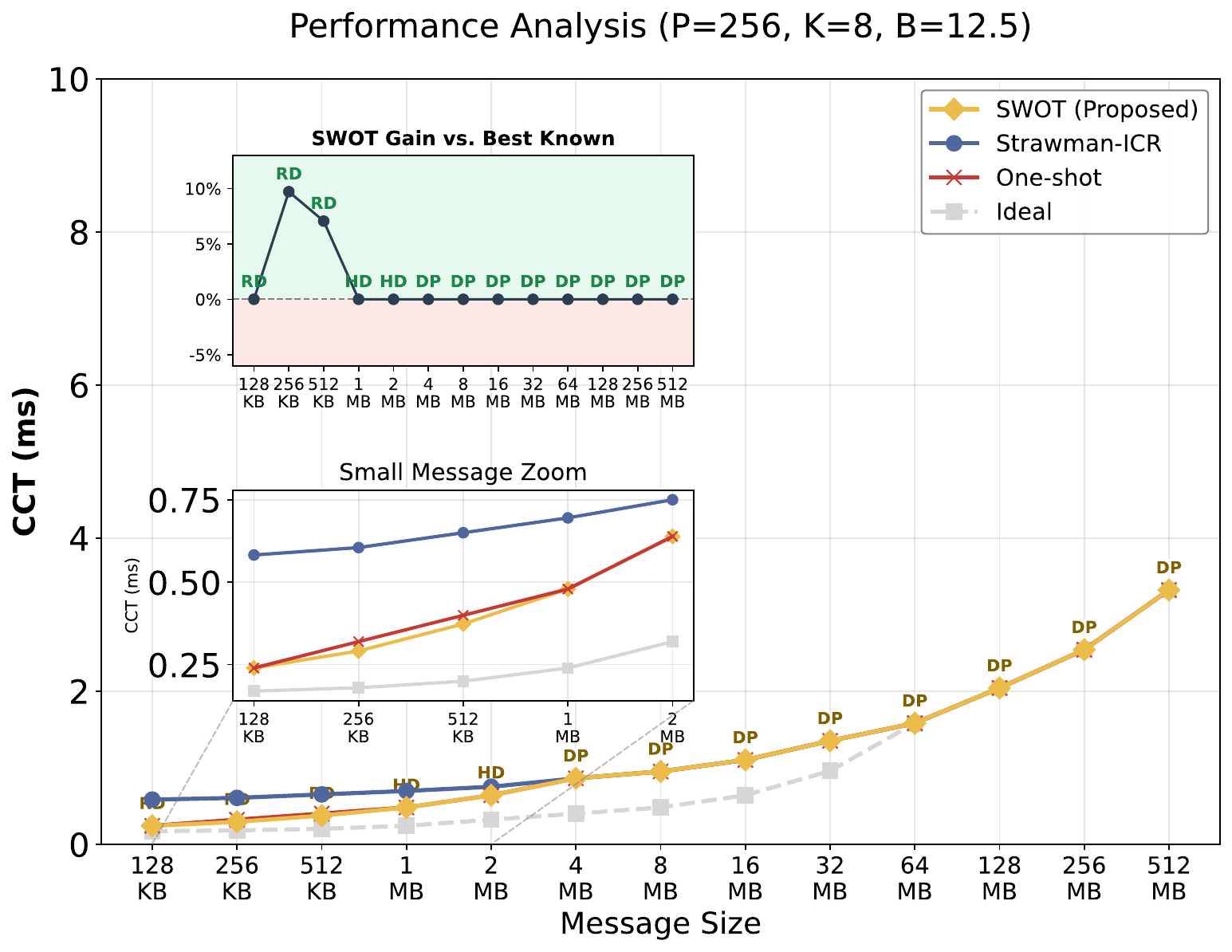}
\label{subfig:primitive-ar}
  }
  \vspace{-0.15in}
  \caption{Normalized CCT for different collective primitives. For each primitive, the results are derived using the best-performing algorithm for the specific configuration. The small plot in the bottom left reports runtime for small messages; top left shows the CCT gain of \ours compared to the combination of best-known algorithms and scheduling strategies. Letters on top denote the abbreviation of the algorithm.}
  \Description{Two normalized performance plots compare collective primitives for All-to-All and AllReduce. Each point uses the best-performing collective algorithm for that configuration, with letter annotations indicating the selected algorithm. The All-to-All plot shows large gains for SWOT at medium message sizes, while the AllReduce plot shows smaller gains because strong static-friendly algorithms already perform close to the best baseline. Insets show small-message runtime and SWOT's relative gain.}
  \label{fig: primitive_cct}
\end{figure*}

\hlmthree{This approach gives each scheduling paradigm access to a strong algorithmic baseline, so smaller gains indicate cases where existing collective algorithms already map well to static optical configurations.}
The results are shown in Figure~\ref{fig: primitive_cct}, where the labels above the data points denote the selected best algorithm (e.g., \textbf{B} for Bruck, \textbf{P} for Pairwise, \textbf{RD} for Recursive Doubling, \textbf{HD} for Rabenseifner's, and \textbf{DP} for Double Binary Tree).
Notably, Double Binary Tree's performance is highly sensitive to pipeline chunk size. We therefore implemented a heuristic chunk selection strategy derived from the NVIDIA NCCL source code to dynamically tune chunk size based on message volume and cluster scale ($p$).

\textbf{All-to-All (Fig. \ref{fig: primitive-a2a}): Dominating Dynamic Patterns.}
For the All-to-All primitive, which exhibits dense and dynamic traffic patterns, \ours demonstrates substantial performance gains.
As shown in the top-left zoom-in plot, \ours achieves a peak speedup of over 93.19\% compared to the best-known baseline at medium message sizes (e.g., 16 MB).
In this range, the \textit{Bruck} algorithm (\textbf{B}) is optimal but requires $\log N$ reconfiguration steps. \ours effectively pipelines these reconfigurations, whereas \strawman incurs heavy switching penalties and one-shot suffers from severe bandwidth fragmentation.
At extremely large message sizes (e.g., 512 MB), the \textit{Pairwise} algorithm (\textbf{P}) becomes dominant. Here, the transmission time significantly exceeds the reconfiguration latency, causing the gap between schemes to narrow as the system becomes bandwidth-bound.

\textbf{AllReduce (Fig. \ref{subfig:primitive-ar}): Near-Optimal Static Baselines.}
For AllReduce, we observe more modest benefits. The traffic patterns of AllReduce (e.g., Ring, Tree) are inherently sparse and map efficiently to static topologies, allowing the one-shot approach to perform near-optimally.
Specifically, \ours achieves a peak improvement of 7.06\%\textendash 9.71\% in the small-to-medium message regime (256 KB\textendash 512 KB) by optimizing the step-wise execution of \textit{Halving-Doubling} (\textbf{HD}).
For larger messages ($> 1$ MB), the one-shot approach remains highly competitive, as the DP algorithm can fully saturate the bandwidth of a static topology without requiring runtime reconfiguration.


\textbf{Insight: The Potential of Co-Design.}
The results above reveal a fundamental limitation: the collective algorithms used in this study (e.g., Bruck, Double Binary Tree) were originally designed under the assumption of a static network.
While \ours extracts performance gains by scheduling these legacy algorithms dynamically, we posit that this is only the tip of the iceberg.
A holistic \textit{co-design} of collective algorithms and reconfigurable topology scheduling--where the communication pattern is inherently designed to exploit temporal reconfiguration--would likely yield significantly larger performance gains. This represents a promising direction for future research.


\subsection{Impact of System Parameters}
In this section, we delve into how physical network parameters dictate the performance gains of \ours. We investigate the influence of optical degree ($k$) and reconfiguration latency ($T_{\text{recfg}}$) to understand the underlying mechanisms of reconfiguration-communication overlapping.

\subsubsection{Impact of Optical Degree ($k$)}
We investigate the impact of the optical degree $k$ (the number of OCS interfaces per node) on \ours's performance. We fix the total bandwidth per node at 800 Gbps and scale $k \in \{2, 4, 8\}$, resulting in per-link bandwidths of 400, 200, and 100 Gbps, respectively. The experiments are conducted on a 256-node cluster with a strict solver time limit of 120 seconds. Figure \ref{fig:impact_k} illustrates the improvement over \strawman. We observe three key trends:

\begin{figure*}[htbp]
    \centering
    \includegraphics[width=\columnwidth]{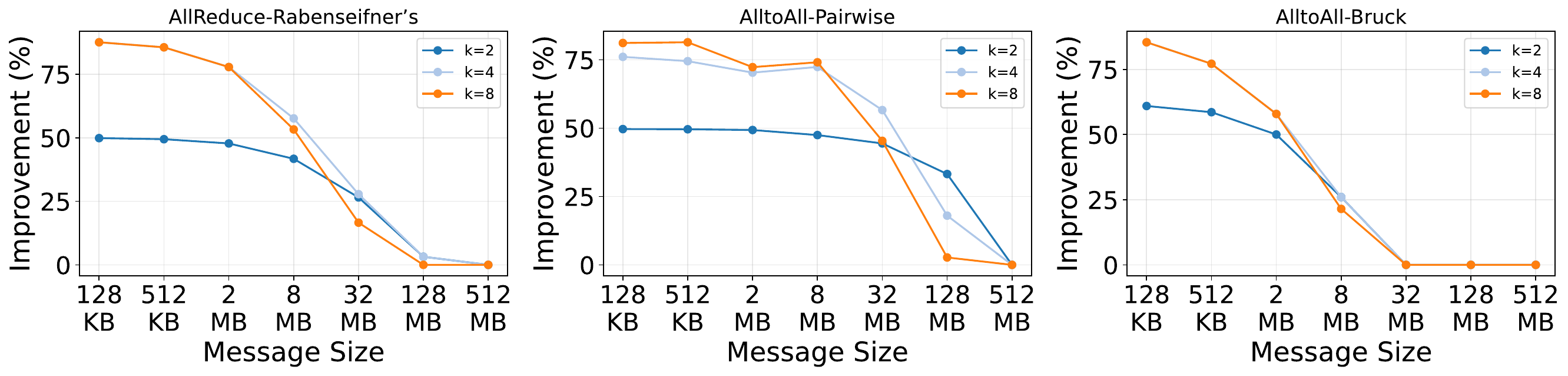}
    \vspace{-0.2in}
    \caption{Impact of optical degree $k$ on Performance Improvement across different algorithms ($p=256$, Solver Limit = 120s). While higher $k$ generally improves performance via finer granularity, the exploded search space at $k=8$ can lead to sub-optimal solutions under strict time constraints.}
    \Description{A set of line plots shows SWOT's performance improvement over Strawman-ICR as message size varies under different optical degrees. Curves for different values of $k$ illustrate that increasing the number of OCS planes often improves scheduling granularity and overlap opportunities, but the $k=8$ curves can flatten or drop in some regimes because the larger optimization search space leads to lower-quality solutions under the solver time limit.}
    \label{fig:impact_k}
\end{figure*}

\textbf{(1) Finer Scheduling Granularity Yields Higher Potential Gains.}
In general, increasing $k$ provides finer scheduling granularity. With more parallel optical lanes, \ours possesses greater flexibility to pipeline reconfiguration and transmission, effectively minimizing the reconfiguration overhead ratio.
\textbf{(2) Diminishing Marginal Returns.}
Increasing $k$ exhibits diminishing returns. As shown in Fig.~\ref{fig:impact_k}, the performance improvement from $k=4$ to $k=8$ is often marginal. This suggests a moderate optical degree (e.g., $k=4$) suffices to exploit most reconfiguration-communication overlap opportunities; further increases yield limited benefit.
\textbf{(3) Trade-off between Search Space and Solution Quality.}
By contrast, we observe instances where a higher $k$ leads to performance degradation. Notably, for the Pairwise and Halving-Doubling algorithms at larger message sizes (e.g., $8$ MB), the performance of $k=8$ drops below that of $k=4$, and sometimes even below $k=2$. This anomaly is attributed to the exponential growth of the solution space.
Under the strict 120-second time limit, the solver may fail to converge to a near-optimal solution for these complex, high-dimensional instances, resulting in sub-optimal schedules.

\subsubsection{Impact of Reconfiguration Latency ($T_{\text{recfg}}$)}
\label{subsec:impact_reconf}
We evaluate the robustness of \ours by varying the reconfiguration latency $T_{\text{recfg}}$ across four orders of magnitude: $\{0.02, 0.2, 2.0, 20.0\}$ ms. This range encompasses diverse optical switching technologies, from ultra-fast switches to slower MEMS-based devices. Experiments are conducted on a 256-node cluster with $k=4$. Figure \ref{fig:impact_latency} presents the reconfiguration overhead ratio across different message sizes. We highlight two observations:

\begin{figure*}[htbp]
    \centering
    \includegraphics[width=\columnwidth]{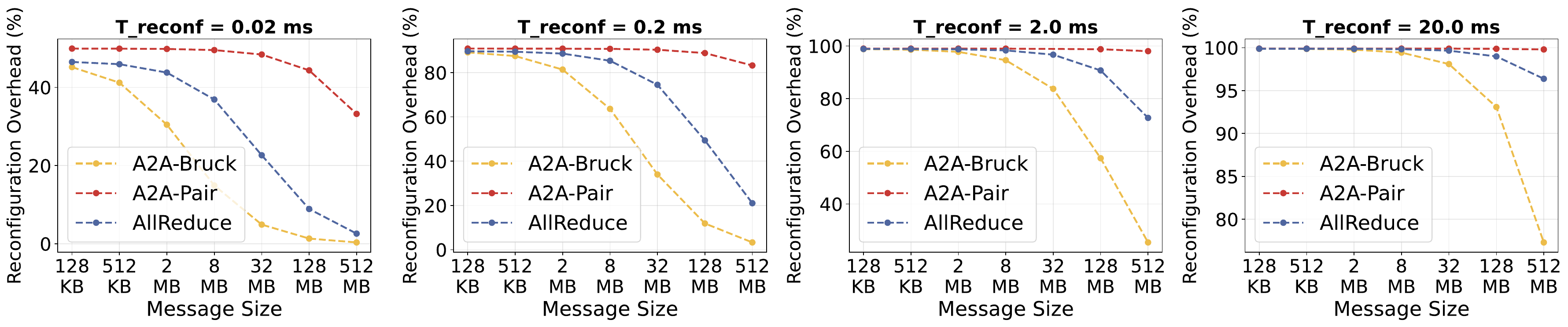}
    \vspace{-0.15in}
    \caption{Reconfiguration Overhead vs. Message Size under varying Reconfiguration Latencies ($T_{\text{recfg}}$). The "knee" of the curve, representing the onset of effective overlapping, shifts to the right as $T_{\text{recfg}}$ increases, confirming that gains are maximized when transmission time exceeds reconfiguration latency.}
    \Description{A line plot shows reconfiguration overhead ratio versus message size for several optical reconfiguration latencies. For short reconfiguration latencies, overhead falls at smaller message sizes; for longer reconfiguration latencies, the drop occurs at larger message sizes. The visible knee of each curve moves right as $T_{\text{recfg}}$ increases, indicating that effective overlap begins when per-step transmission time becomes large enough to hide reconfiguration delay.}
    \label{fig:impact_latency}
\end{figure*}

\textbf{(1) Latency Robustness across Switching Scales.}
\ours demonstrates adaptability across varying $T_{\text{recfg}}$ in our schedule-level model.
While absolute overhead naturally increases with slower switches (e.g., $T_{\text{recfg}}=20$ ms), the framework consistently identifies valid schedules; larger latency mainly shifts the effective overlap regime toward larger per-step transmissions.
This indicates that \ours can tolerate a broad range of conservative path-readiness latencies without relying on ultra-fast switching alone.

\textbf{(2) The Mechanics of Overlapping: The Critical Ratio.}
The results reveal that the efficacy of the overlap technique is fundamentally governed by the ratio between the per-step transmission duration ($t_{\text{trans}} = m_{step}/B$) and the reconfiguration latency ($T_{\text{recfg}}$). We observe a "critical inflection point" in the overhead curves:
\begin{itemize}
    \item \textbf{Latency-Dominated Regime} ($t_{\text{trans}} < T_{\text{recfg}}$): When message size is small, transmission time is insufficient to cover the reconfiguration delay of the subsequent step. In this regime (the flat plateaus on the left of Fig.~\ref{fig:impact_latency}), the system gain from overlapping is minimal, as the bottleneck is strictly the switching speed.
    \item \textbf{Bandwidth-Dominated Regime}($t_{\text{trans}} \ge T_{\text{recfg}}$): As message size increases, $t_{\text{trans}}$ exceeds $T_{\text{recfg}}$. The scheduler effectively overlaps OCS reconfiguration latency with ongoing transmissions on other OCSs. As a result, the reconfiguration overhead ratio is greatly reduced.
\end{itemize}
As $T_{\text{recfg}}$ increases from 0.2~ms to 20~ms, the inflection point shifts to the right, requiring larger message sizes to achieve effective overlap.
Moreover, the improvement rate is algorithm-dependent.
Algorithms with larger per-step data chunks (e.g., A2A-Bruck) enter the bandwidth-dominated regime earlier, whereas those with more fragmented traffic patterns (e.g., A2A-Pairwise) require substantially larger messages to generate sufficient transmission time to mask the latency.


This result highlights a key advantage of \ours: it relaxes the stringent requirement for ultra-low latency optical switches.
By creating overlap windows, \ours potentially enables high-performance collective communication on commodity OCS hardware when per-step transmissions are long enough to cover a meaningful fraction of path-readiness latency.

\section{Related Work}

\textbf{\hlmthree{Optical Interconnects for General-Purpose Datacenters.}}
\hlmthree{General-purpose optical datacenter interconnects are often categorized by how much traffic information they use.
Demand-aware designs such as Helios~\cite{farringtonHeliosHybridElectrical2010} and c-Through~\cite{wangCThroughParttime2010} measure traffic demand and configure optical circuits for high-volume flows, whereas demand-oblivious designs such as RotorNet~\cite{melletteRotorNetScalableLowcomplexity2017} rotate through a predetermined circuit schedule without reacting to instantaneous traffic.}

\textbf{Optical Interconnects for ML.}
Optical switching has been widely explored to accelerate DML.
Systems like SiP-ML~\cite{khaniSiPMLHighbandwidthOptical2021}, TopoOpt~\cite{wangTopoOptCooptimizingNetwork2023}, \hlmthree{InfiniteHBD~\cite{shouInfiniteHBDBuildingDatacenterScale2025}}, and Google TPUv4~\cite{jouppiTPUV4Optically2023,zuResiliencyScaleManaging2024a} leverage OCS to build demand-aware topologies. However, they predominantly adopt a ``one-shot'' strategy, fixing the topology for entire training iterations to avoid reconfiguration overhead.
Recent works like ACTINA~\cite{wuACTINAAdaptingCircuitSwitching2025} and MixNet~\cite{liaoMixNetRuntimeReconfigurable2025} introduce dynamism by reconfiguring between computation phases, yet they still treat collective operations as atomic, static blocks.
Crucially, while Athapathu et al.~\cite{athapathuReconfigurabilityCollectiveCommunication2025} theoretically analyzed intra-collective reconfiguration, they concluded that performance gains are restricted to ultra-fast switches ($<500$ ns).
\ours challenges this conclusion, demonstrating that by overlapping reconfiguration with transmission, significant gains are achievable even with commodity microsecond-scale switches.

\textbf{Collective Algorithm Synthesis.}
Significant research focuses on the design and synthesis of efficient CC algorithms for specific topologies. SCCL~\cite{caiSynthesizingOptimalCollective2021} pioneered a systematic approach to synthesizing collective algorithms. Subsequent works explored diverse optimization strategies:
TACCL~\cite{shahTACCLGuidingCollective2023} improves scalability via communication sketches but trades off schedule quality;
TE-CCL~\cite{xutingliuRethinkingMachineLearning2024} adopts a traffic-engineering-based formulation;
SyCCL~\cite{caoSyCCLExploitingSymmetry2025} exploits collective symmetries for efficient and accurate schedule synthesis.
Notably, Zhao et al.  \cite{zhaoEfficientDirectConnectTopologies2025} jointly optimize static direct-connect topologies and their corresponding collective schedules at scale.
\hlmthree{\ours is complementary to this line of work: it does not synthesize new CC algorithms, but schedules the execution of a selected algorithm on reconfigurable optical planes.}

\textbf{\hlmthree{Traffic Routing and Matrix Decomposition.}}
\hlmthree{Traffic-routing and Birkhoff-von Neumann-style methods decompose or route demand matrices in reconfigurable datacenter networks to avoid congestion~\cite{vallsBirkhoffsDecompositionRevisited2020,zerwasD3AdaptiveReconfigurable2024}.
This objective is orthogonal to \ours: each selected collective step in our input schedule already specifies congestion-free pairwise transfers, while \ours focuses on coordinating OCS reconfiguration with step dependencies and data transmission.}

\section{Discussion}
\label{sec:discussion}

\textbf{\hlmtwo{Applicability Scope.}}
\hlmtwo{\ours targets recurrent collective communication whose demand can be represented as an input schedule before execution, either from the selected collective algorithm or from profiling across iterations.
This assumption is common in dense-model training, where communication patterns are largely determined by model architecture, collective algorithm, and parallelization strategy~\cite{liUnderstandingCommunicationCharacteristics2024,liAnalyzingCommunicationPredictability2025}.}
\hlmtwo{Our evaluation instantiates this setting on the direct-connect optical substrate introduced in \S\ref{sec:background_motivation}; we first consider how this scope relates to routed multi-dimensional optical fabrics and dynamic sparse workloads.}

\textbf{\hlmtwo{Multi-dimensional Optical Topologies.}}
\hlmtwo{Multi-dimensional optical fabrics are constructed differently.
In systems such as TPUv4, chips are organized into 3D torus or twisted-torus topologies; OCSes configure inter-cube or dimension-specific ICI links rather than providing a single OCS plane that connects every node to every other node~\cite{jouppiTPUV4Optically2023,zuResiliencyScaleManaging2024a}.
Therefore, a source-destination pair that is one optical hop in our flattened model may require several dimension-local hops in a 2D/3D topology.
The topology itself determines the route, and an OCS configuration controls only the relevant dimension-local link set.}
\hlmtwo{This changes the scheduler input.
Instead of mapping a collective step directly to one global permutation $\text{cfg}_i$, a multi-dimensional extension would first decompose each logical collective transfer into route-segment tasks, each attached to a dimension, local OCS resource, capacity, and dependency.
After this decomposition, the core \ours decisions remain meaningful: assign traffic to schedulable optical resources, overlap reconfiguration with transfers on other resources, and bypass configurations that are not needed by a route segment.}
\hlmtwo{At the execution level, the same guarded-operation contract would apply to route-segment tasks rather than to a single global OCS plane.
Developing this multi-dimensional version would require a routing/decomposition layer, the corresponding multi-dimensional MILP constraints, and a fair evaluation against optimized multi-hop collectives, which remain outside this paper.}


\textbf{\hlmtwo{Dynamic Sparse Workloads.}}
\hlmtwo{The key requirement above is that the communication demand can be represented as an input schedule before the offline planner runs.
Thus, workloads whose communication is generated by a known collective algorithm or by a stable profiled pattern can be handled by the same framework.}
\hlmtwo{MoE-style sparse training is more challenging because expert routing can make the traffic matrix input-dependent.
Recent studies show that dense LLM traffic is often predictable from model and parallelization choices, while MoE traffic may still expose semi-predictability or locality that can be exploited by the network layer~\cite{liAnalyzingCommunicationPredictability2025,liaoMixNetRuntimeReconfigurable2025}.}
\hlmtwo{When such sparse traffic can be predicted or grouped into reusable schedules, \ours can operate at the framework level without changing its execution principle.
Fully dynamic traffic would require additional mechanisms such as prediction, schedule banks, locality-aware domain selection, fallback schedules, or online replanning, which we leave to future work.}

\textbf{Algorithmic Co-design Opportunities.}
\ours currently acts as an accelerator for existing CC algorithms designed for static networks.
\hlmthree{The per-algorithm results show that this approach can benefit several legacy algorithms.
At the same time, the apples-to-apples comparison against best-known algorithms suggests that the remaining end-to-end gains can be more limited, particularly for AllReduce primitives.}
\hlmthree{One contributing factor is that these best-known CC algorithms were designed under static or fixed-topology network assumptions.
As a result, they may not naturally expose communication steps with enough topology variation, transmission slack, or bypass opportunities for a reconfigurable optical fabric to fully exploit.}
\hlmthree{This points to a broader opportunity: co-designing CC algorithms with topology reconfiguration, so that communication pairings, step structures, and reconfiguration opportunities are considered together.
Within this design space, \ours's scheduling techniques could further exploit overlap and topology bypassing, potentially yielding larger gains than accelerating unchanged static-network algorithms.}
\hlmthree{We view this as an important algorithm-network co-design opportunity and a promising future direction.} 



\section{Conclusion}
In this paper, we present \ours, an \newapproachst reconfigurable optical framework that dynamically aligns network resources with the communication demands of individual CC algorithms.
\ours strategicall splits messages, selects target optical topologies, and schedules optical reconfigurations asynchronously to overlap reconfiguration with data transmission.
By hiding optical reconfiguration costs, \ours reduces topology switching overhead while remaining compatible with existing collective libraries.
Our evaluation demonstrates up to 89.7\% communication time reduction across diverse primitives, with robust performance against varying optical degrees and reconfiguration latencies spanning four orders of magnitude.
This work introduces a co-adaptive paradigm between optical networks and dynamic DML communication, offering a potential pathway toward scalable infrastructure for future AI training systems.

\textbf{Ethics.} This work does not raise ethical concerns.

\begin{acks}
We thank the anonymous reviewers and our shepherd Pooria Namyar for their constructive feedback.
This work is supported by the National Natural Science Foundation of China (NSFC) under Grant 62372426, the Youth Innovation Promotion Association of the Chinese Academy of Science under Grant 2023481, Basic Research Program of Jiangsu under Grant BK20250124, and the Fundamental Research Funds for the Central Universities under Grant WK2150250043.
\end{acks}

\bibliographystyle{ACM-Reference-Format}
\bibliography{main}

@misc{kaplanScalingLawsNeural2020,
  author =        {Kaplan, Jared and McCandlish, Sam and Henighan, Tom and
                   Brown, Tom B. and Chess, Benjamin and Child, Rewon and
                   Gray, Scott and Radford, Alec and Wu, Jeffrey and
                   Amodei, Dario},
  number =        {arXiv:2001.08361},
  publisher =     {arXiv},
  title =         {Scaling {{Laws}} for {{Neural Language Models}}},
  year =          {2020},
}

@inproceedings{jiangMegaScaleScalingLarge2024a,
  address =       {Santa Clara, CA, USA},
  author =        {Jiang, Ziheng and Lin, Haibin and Zhong, Yinmin and
                   Huang, Qi and Chen, Yangrui and Zhang, Zhi and
                   Peng, Yanghua and Li, Xiang and Xie, Cong and
                   Nong, Shibiao and Jia, Yulu and He, Sun and
                   Chen, Hongmin and Bai, Zhihao and Hou, Qi and
                   Yan, Shipeng and Zhou, Ding and Sheng, Yiyao and
                   Jiang, Zhuo and Xu, Haohan and Wei, Haoran and
                   Zhang, Zhang and Nie, Pengfei and Zou, Leqi and
                   Zhao, Sida and Xiang, Liang and Liu, Zherui and
                   Li, Zhe and Jia, Xiaoying and Ye, Jianxi and Jin, Xin and
                   Liu, Xin},
  booktitle =     {21st {{USENIX Symposium}} on {{Networked Systems
                   Design}} and {{Implementation}} ({{NSDI}} 24)},
  pages =         {745--760},
  publisher =     {USENIX Association},
  title =         {{{MegaScale}: Scaling Large Language Model Training to More
                   Than 10,000 {GPUs}}},
  year =          {2024},
  isbn =          {978-1-939133-39-7},
}

@inproceedings{wangDomainSpecificNetworkTransport2024,
  address =       {Santa Clara, CA, USA},
  author =        {Wang, Hao and Tian, Han and Chen, Jingrong and
                   Wan, Xinchen and Xia, Jiacheng and Zeng, Gaoxiong and
                   Bai, Wei and Jiang, Junchen and Wang, Yong and
                   Chen, Kai},
  booktitle =     {21st {{USENIX Symposium}} on {{Networked Systems
                   Design}} and {{Implementation}} ({{NSDI}} 24)},
  pages =         {1421--1443},
  publisher =     {USENIX Association},
  title =         {Towards {Domain-Specific} Network Transport for Distributed
                   {DNN} Training},
  year =          {2024},
  isbn =          {978-1-939133-39-7},
}

@inproceedings{liUnderstandingCommunicationCharacteristics2024,
  address =       {New York, NY, USA},
  author =        {Li, Wenxue and Liu, Xiangzhou and Li, Yuxuan and
                   Jin, Yilun and Tian, Han and Zhong, Zhizhen and
                   Liu, Guyue and Zhang, Ying and Chen, Kai},
  booktitle =     {Proceedings of the 8th {{Asia-Pacific Workshop}} on
                   {{Networking}}},
  pages =         {1--8},
  publisher =     {Association for Computing Machinery},
  series =        {{{APNet}} '24},
  title =         {Understanding Communication Characteristics of Distributed
                   Training},
  year =          {2024},
  isbn =          {979-8-4007-1758-1},
}

@misc{liAnalyzingCommunicationPredictability2025,
      title={Analyzing Communication Predictability in LLM Training},
      author={Wenxue Li and Xiangzhou Liu and Yuxuan Li and Yilun Jin and Zhenghang Ren and Xudong Liao and Han Tian and Bo Ren and Zhizhen Zhong and Guyue Liu and Ying Zhang and Kai Chen},
      year={2025},
      eprint={2512.24750},
      archivePrefix={arXiv},
      primaryClass={cs.NI},
      url={https://arxiv.org/abs/2512.24750},
}

@inproceedings{jouppiTPUV4Optically2023,
  address =       {New York, NY, USA},
  author =        {Jouppi, Norm and Kurian, George and Li, Sheng and
                   Ma, Peter and Nagarajan, Rahul and Nai, Lifeng and
                   Patil, Nishant and Subramanian, Suvinay and
                   Swing, Andy and Towles, Brian and Young, Clifford and
                   Zhou, Xiang and Zhou, Zongwei and Patterson, David A},
  booktitle =     {Proceedings of the 50th {{Annual International
                   Symposium}} on {{Computer Architecture}}},
  pages =         {1--14},
  publisher =     {Association for Computing Machinery},
  series =        {{{ISCA}} '23},
  title =         {{{TPU}} v4: {{An Optically Reconfigurable
                   Supercomputer}} for {{Machine Learning}} with
                   {{Hardware Support}} for {{Embeddings}}},
  year =          {2023},
  isbn =          {9798400700958},
}

@inproceedings{zuResiliencyScaleManaging2024a,
  address =       {Santa Clara, CA, USA},
  author =        {Zu, Yazhou and Ghaffarkhah, Alireza and
                   Dang, Hoang-Vu and Towles, Brian and Hand, Steven and
                   Huda, Safeen and Bello, Adekunle and
                   Kolbasov, Alexander and Rezaei, Arash and Du, Dayou and
                   Lacy, Steve and Wang, Hang and Wisner, Aaron and
                   Lewis, Chris and Bahini, Henri},
  booktitle =     {21st {{USENIX Symposium}} on {{Networked Systems
                   Design}} and {{Implementation}} ({{NSDI}} 24)},
  pages =         {761--774},
  publisher =     {USENIX Association},
  title =         {Resiliency at Scale: Managing {Google}'s {TPUv4} Machine
                   Learning Supercomputer},
  year =          {2024},
  isbn =          {978-1-939133-39-7},
}

@inproceedings{poutievskiJupiterEvolvingTransforming2022,
  address =       {New York, NY, USA},
  author =        {Leon and Mashayekhi, Omid and Ong, Joon and
                   Singh, Arjun and Tariq, Mukarram and Wang, Rui and
                   Zhang, Jianan and Beauregard, Virginia and
                   Conner, Patrick and Gribble, Steve and Kapoor, Rishi and
                   Kratzer, Stephen and Li, Nanfang and Liu, Hong and
                   Nagaraj, Karthik and Ornstein, Jason and
                   Sawhney, Samir and Urata, Ryohei and
                   Vicisano, Lorenzo and Yasumura, Kevin and
                   Zhang, Shidong and Zhou, Junlan and Vahdat, Amin},
  booktitle =     {Proceedings of the {{ACM SIGCOMM}} 2022
                   {{Conference}}},
  pages =         {66--85},
  publisher =     {Association for Computing Machinery},
  series =        {{{SIGCOMM}} '22},
  title =         {Jupiter Evolving: Transforming Google's Datacenter
                   Network via Optical Circuit Switches and
                   Software-Defined Networking},
  year =          {2022},
  isbn =          {978-1-4503-9420-8},
}

@inproceedings{liuLightwaveFabricsAtScale2023,
  address =       {New York, NY, USA},
  author =        {Liu, Hong and Urata, Ryohei and Yasumura, Kevin and
                   Zhou, Xiang and Bannon, Roy and Berger, Jill and
                   Dashti, Pedram and Jouppi, Norm and Lam, Cedric and
                   Li, Sheng and Mao, Erji and Nelson, Daniel and
                   Papen, George and Tariq, Mukarram and Vahdat, Amin},
  booktitle =     {Proceedings of the {{ACM SIGCOMM}} 2023
                   {{Conference}}},
  pages =         {499--515},
  publisher =     {Association for Computing Machinery},
  series =        {{{ACM SIGCOMM}} '23},
  title =         {Lightwave {{Fabrics}}: {{At-Scale Optical Circuit
                   Switching}} for {{Datacenter}} and {{Machine Learning
                   Systems}}},
  year =          {2023},
  isbn =          {9798400702365},
}

@inproceedings{ghobadiEmergingOpticalInterconnects2022,
  address =       {San Diego, CA, USA},
  author =        {Ghobadi, Manya},
  booktitle =     {Optical Fiber Communication Conference ({{OFC}})
                   2022},
  pages =         {Th1G.1},
  publisher =     {Optica Publishing Group},
  series =        {Optical {{Fiber Communication Conference}} ({{OFC}})
                   2022},
  title =         {Emerging Optical Interconnects for {{AI}} Systems},
  year =          {2022},
}

@inproceedings{khaniSiPMLHighbandwidthOptical2021,
  address =       {New York, NY, USA},
  author =        {Khani, Mehrdad and Ghobadi, Manya and
                   Alizadeh, Mohammad and Zhu, Ziyi and Glick, Madeleine and
                   Bergman, Keren and Vahdat, Amin and Klenk, Benjamin and
                   Ebrahimi, Eiman},
  booktitle =     {Proceedings of the 2021 {{ACM SIGCOMM}} 2021
                   {{Conference}}},
  pages =         {657--675},
  publisher =     {Association for Computing Machinery},
  series =        {{{SIGCOMM}} '21},
  title =         {{{SiP-ML}}: High-Bandwidth Optical Network
                   Interconnects for Machine Learning Training},
  year =          {2021},
  isbn =          {978-1-4503-8383-7},
}

@inproceedings{wangTopoOptCooptimizingNetwork2023,
  address =       {Boston, MA, USA},
  author =        {Wang, Weiyang and Khazraee, Moein and Zhong, Zhizhen and
                   Ghobadi, Manya and Jia, Zhihao and
                   Mudigere, Dheevatsa and Zhang, Ying and
                   Kewitsch, Anthony},
  booktitle =     {20th {{USENIX Symposium}} on {{Networked Systems
                   Design}} and {{Implementation}} ({{NSDI}} 23)},
  pages =         {739--767},
  publisher =     {USENIX Association},
  series =        {{{NSDI}}'23},
  title =         {{{TopoOpt}: Co-optimizing Network Topology and
                   Parallelization Strategy for Distributed Training Jobs}},
  year =          {2023},
  isbn =          {978-1-939133-33-5},
}

@misc{shouInfiniteHBDBuildingDatacenterScale2025,
  author =        {Shou, Chenchen and Liu, Guyue and Nie, Hao and Meng, Huaiyu and
                   Zhou, Yu and Jiang, Yimin and Lv, Wenqing and Xu, Yelong and
                   Lu, Yuanwei and Chen, Zhang and Yu, Yanbo and Shen, Yichen and
                   Zhu, Yibo and Jiang, Daxin},
  title =         {{{InfiniteHBD}}: Building Datacenter-Scale High-Bandwidth
                   Domain for {{LLM}} with Optical Circuit Switching
                   Transceivers},
  year =          {2025},
  eprint =        {2502.03885},
  archiveprefix = {arXiv},
  primaryclass =  {cs.NI},
  doi =           {10.48550/arXiv.2502.03885},
}

@inproceedings{farringtonHeliosHybridElectrical2010,
  address =       {New York, NY, USA},
  author =        {Farrington, Nathan and Porter, George and
                   Radhakrishnan, Sivasankar and Bazzaz, Hamid Hajabdolali and
                   Subramanya, Vikram and Fainman, Yeshaiahu and Papen, George and
                   Vahdat, Amin},
  booktitle =     {Proceedings of the {{ACM SIGCOMM}} 2010 {{Conference}}},
  doi =           {10.1145/1851182.1851223},
  pages =         {339--350},
  publisher =     {Association for Computing Machinery},
  series =        {{{SIGCOMM}} '10},
  title =         {Helios: {{A}} Hybrid Electrical/Optical Switch Architecture
                   for Modular Data Centers},
  year =          {2010},
  isbn =          {978-1-4503-0201-2},
}

@inproceedings{wangCThroughParttime2010,
  address =       {New York, NY, USA},
  author =        {Wang, Guohui and Andersen, David G. and Kaminsky, Michael and
                   Papagiannaki, Konstantina and Ng, T. S. Eugene and
                   Kozuch, Michael and Ryan, Michael},
  booktitle =     {Proceedings of the {{ACM SIGCOMM}} 2010 {{Conference}}},
  doi =           {10.1145/1851182.1851222},
  pages =         {327--338},
  publisher =     {Association for Computing Machinery},
  series =        {{{SIGCOMM}} '10},
  title =         {c-Through: Part-Time Optics in Data Centers},
  year =          {2010},
  isbn =          {978-1-4503-0201-2},
}

@inproceedings{ballaniSiriusFlatDatacenter2020,
  address =       {New York, NY, USA},
  author =        {Ballani, Hitesh and Costa, Paolo and
                   Behrendt, Raphael and Cletheroe, Daniel and
                   Haller, Istvan and Jozwik, Krzysztof and
                   Karinou, Fotini and Lange, Sophie and Shi, Kai and
                   Thomsen, Benn and Williams, Hugh},
  booktitle =     {Proceedings of the {{Annual}} Conference of the {{ACM
                   Special Interest Group}} on {{Data Communication}} on
                   the Applications, Technologies, Architectures, and
                   Protocols for Computer Communication},
  pages =         {782--797},
  publisher =     {Association for Computing Machinery},
  series =        {{{SIGCOMM}} '20},
  title =         {Sirius: {{A Flat Datacenter Network}} with
                   {{Nanosecond Optical Switching}}},
  year =          {2020},
  isbn =          {978-1-4503-7955-7},
}

@inproceedings{melletteRotorNetScalableLowcomplexity2017,
  address =       {New York, NY, USA},
  author =        {Mellette, William M. and McGuinness, Rob and
                   Roy, Arjun and Forencich, Alex and Papen, George and
                   Snoeren, Alex C. and Porter, George},
  booktitle =     {Proceedings of the {{Conference}} of the {{ACM
                   Special Interest Group}} on {{Data Communication}}},
  pages =         {267--280},
  publisher =     {Association for Computing Machinery},
  series =        {{{SIGCOMM}} '17},
  title =         {{{RotorNet}}: {{A Scalable}}, {{Low-complexity}},
                   {{Optical Datacenter Network}}},
  year =          {2017},
  isbn =          {978-1-4503-4653-5},
}

@inproceedings{gengExploitingNaturalNetwork2018,
  address =       {Renton, WA},
  author =        {Geng, Yilong and Liu, Shiyu and Yin, Zi and
                   Naik, Ashish and Prabhakar, Balaji and
                   Rosenblum, Mendel and Vahdat, Amin},
  booktitle =     {15th {{USENIX Symposium}} on {{Networked Systems
                   Design}} and {{Implementation}} ({{NSDI}} 18)},
  pages =         {81--94},
  publisher =     {USENIX Association},
  title =         {Exploiting a Natural Network Effect for Scalable,
                   Fine-grained Clock Synchronization},
  year =          {2018},
  isbn =          {978-1-939133-01-4},
  url =           {https://www.usenix.org/conference/nsdi18/presentation/geng},
}

@inproceedings{namyarFireflyScalableUltraAccurate2025,
  address =       {New York, NY, USA},
  author =        {Namyar, Pooria and Li, Yuliang and Wang, Weitao and
                   Dukkipati, Nandita and Yap, {KK} and Gong, Junzhi and
                   Chen, Chen and Gao, Peixuan and Ray, Devdeep and
                   Kumar, Gautam and Ma, Yidan and Govindan, Ramesh and
                   Vahdat, Amin},
  booktitle =     {Proceedings of the {{ACM SIGCOMM}} 2025 {{Conference}}},
  doi =           {10.1145/3718958.3750502},
  pages =         {434--452},
  publisher =     {Association for Computing Machinery},
  series =        {{{SIGCOMM}} '25},
  title =         {Firefly: Scalable, Ultra-Accurate Clock Synchronization
                   for Datacenters},
  year =          {2025},
  isbn =          {979-8-4007-1524-2},
}

@misc{zerwasD3AdaptiveReconfigurable2024,
  author =        {Zerwas, Johannes and Griner, Chen and Schmid, Stefan and
                   Avin, Chen},
  title =         {{{D3}}: An Adaptive Reconfigurable Datacenter Network},
  year =          {2024},
  eprint =        {2406.13380},
  archiveprefix = {arXiv},
  primaryclass =  {cs.NI},
  doi =           {10.48550/arXiv.2406.13380},
}

@misc{vallsBirkhoffsDecompositionRevisited2020,
  author =        {Valls, Victor and Iosifidis, George and Tassiulas, Leandros},
  title =         {Birkhoff's Decomposition Revisited: Sparse Scheduling for
                   High-Speed Circuit Switches},
  year =          {2020},
  eprint =        {2011.02752},
  archiveprefix = {arXiv},
  primaryclass =  {math.OC},
  doi =           {10.48550/arXiv.2011.02752},
}

@article{yeDeepLearningWorkload2024,
  author =        {Ye, Zhisheng and Gao, Wei and Hu, Qinghao and
                   Sun, Peng and Wang, Xiaolin and Luo, Yingwei and
                   Zhang, Tianwei and Wen, Yonggang},
  journal =       {ACM Computing Surveys},
  number =        {6},
  pages =         {146:1--146:38},
  title =         {Deep {{Learning Workload Scheduling}} in {{GPU
                   Datacenters}}: {{A Survey}}},
  volume =        {56},
  year =          {2024},
  issn =          {0360-0300},
}

@incollection{rabenseifnerOptimizationCollectiveReduction2004,
  address =       {Berlin, Heidelberg},
  author =        {Rabenseifner, Rolf},
  booktitle =     {Computational {{Science}} - {{ICCS}} 2004},
  editor =        {Kanade, Takeo and Kittler, Josef and
                   Kleinberg, Jon M. and Mattern, Friedemann and
                   Mitchell, John C. and Naor, Moni and
                   Nierstrasz, Oscar and Pandu Rangan, C. and
                   Steffen, Bernhard and Sudan, Madhu and
                   Terzopoulos, Demetri and Tygar, Dough and
                   Vardi, Moshe Y. and Weikum, Gerhard and Bubak, Marian and
                   Van Albada, Geert Dick and Sloot, Peter M. A. and
                   Dongarra, Jack},
  pages =         {1--9},
  publisher =     {Springer Berlin Heidelberg},
  title =         {Optimization of {{Collective Reduction Operations}}},
  volume =        {3036},
  year =          {2004},
  isbn =          {978-3-540-22114-2 978-3-540-24685-5},
}

@article{cococcioniBigMMethodNumerical2021,
  author =        {Cococcioni, Marco and Fiaschi, Lorenzo},
  journal =       {Optimization Letters},
  number =        {7},
  pages =         {2455--2468},
  title =         {The {{Big-M}} Method with the Numerical Infinite
                   {{M}}},
  volume =        {15},
  year =          {2021},
  issn =          {1862-4480},
}

@article{bruckEfficientAlgorithmsAlltoall1997,
  author =        {Bruck, J. and Ho, Ching-Tien and Kipnis, S. and
                   Upfal, E. and Weathersby, D.},
  journal =       {IEEE Transactions on Parallel and Distributed
                   Systems},
  number =        {11},
  pages =         {1143--1156},
  title =         {Efficient Algorithms for All-to-All Communications in
                   Multiport Message-Passing Systems},
  volume =        {8},
  year =          {1997},
  issn =          {1558-2183},
}

@inproceedings{athapathuReconfigurabilityCollectiveCommunication2025,
  title = {Reconfigurability within {{Collective Communication Algorithms}}},
  booktitle = {Proceedings of the 2nd {{Workshop}} on {{Networks}} for {{AI Computing}}},
  author = {Athapathu, Rukshani and Porter, George},
  year = {2025},
  series = {{{NAIC}} '25},
  pages = {43--49},
  publisher = {Association for Computing Machinery},
  address = {New York, NY, USA},
  urldate = {2025-09-09},
  isbn = {979-8-4007-2082-6}
}

@inproceedings{liaoMixNetRuntimeReconfigurable2025,
  title = {{{MixNet}}: {{A Runtime Reconfigurable Optical-Electrical Fabric}} for {{Distributed Mixture-of-Experts Training}}},
  shorttitle = {{{MixNet}}},
  booktitle = {Proceedings of the {{ACM SIGCOMM}} 2025 {{Conference}}},
  author = {Liao, Xudong and Sun, Yijun and Tian, Han and Wan, Xinchen and Jin, Yilun and Wang, Zilong and Ren, Zhenghang and Huang, Xinyang and Li, Wenxue and Tse, Kin Fai and Zhong, Zhizhen and Liu, Guyue and Zhang, Ying and Ye, Xiaofeng and Zhang, Yiming and Chen, Kai},
  year = {2025},
  series = {{{SIGCOMM}} '25},
  pages = {554--574},
  publisher = {Association for Computing Machinery},
  address = {New York, NY, USA},
  urldate = {2025-09-08},
  isbn = {979-8-4007-1524-2}
}

@misc{xueOpticalSwitchingData2023,
  title = {Optical {{Switching Data Center Networks}}: {{Understanding Techniques}} and {{Challenges}}},
  shorttitle = {Optical {{Switching Data Center Networks}}},
  author = {Xue, Xuwei and Zhang, Shaojuan and Guo, Bingli and Ji, Wei and Yin, Rui and Chen, Bin and Huang, Shanguo},
  year = {2023},
  number = {arXiv:2302.05298},
  eprint = {2302.05298},
  primaryclass = {cs, eess},
  publisher = {arXiv},
  urldate = {2024-03-12},
  archiveprefix = {arXiv}
}

@inproceedings{zhaoEfficientDirectConnectTopologies2025,
  address = {Philadelphia, PA, USA},
  title = {Efficient {Direct-Connect} Topologies for Collective Communications},
  booktitle = {22nd {{USENIX Symposium}} on {{Networked Systems Design}} and {{Implementation}} ({{NSDI}} 25)},
  author = {Zhao, Liangyu and Pal, Siddharth and Chugh, Tapan and Wang, Weiyang and Fantl, Jason and Basu, Prithwish and Khoury, Joud and Krishnamurthy, Arvind},
  year = {2025},
  pages = {705--737},
  publisher = {USENIX Association},
  urldate = {2025-05-07},
  isbn = {978-1-939133-46-5}
}

@inproceedings{caiSynthesizingOptimalCollective2021,
  title = {Synthesizing Optimal Collective Algorithms},
  shorttitle = {{{SCCL}}},
  booktitle = {Proceedings of the 26th {{ACM SIGPLAN Symposium}} on {{Principles}} and {{Practice}} of {{Parallel Programming}}},
  author = {Cai, Zixian and Liu, Zhengyang and Maleki, Saeed and Musuvathi, Madanlal and Mytkowicz, Todd and Nelson, Jacob and Saarikivi, Olli},
  year = 2021,
  series = {{{PPoPP}} '21},
  pages = {62--75},
  publisher = {Association for Computing Machinery},
  address = {New York, NY, USA},
  urldate = {2024-08-29},
  isbn = {978-1-4503-8294-6}
}

@inproceedings{caoSyCCLExploitingSymmetry2025,
  title = {{{SyCCL}}: {{Exploiting Symmetry}} for {{Efficient Collective Communication Scheduling}}},
  shorttitle = {{{SyCCL}}},
  booktitle = {Proceedings of the {{ACM SIGCOMM}} 2025 {{Conference}}},
  author = {Cao, Jiamin and Shi, Shangfeng and Gao, Jiaqi and Liu, Weisen and Yang, Yifan and Xu, Yichi and Zheng, Zhilong and Guan, Yu and Qian, Kun and Liu, Ying and Xu, Mingwei and Wang, Tianshu and Wang, Ning and Dong, Jianbo and Fu, Binzhang and Cai, Dennis and Zhai, Ennan},
  year = 2025,
  series = {{{SIGCOMM}} '25},
  pages = {645--662},
  publisher = {Association for Computing Machinery},
  address = {New York, NY, USA},
  urldate = {2025-09-03},
  isbn = {979-8-4007-1524-2}
}

@inproceedings{sensiSwingShortcuttingRings2024,
  address = {Santa Clara, CA, USA},
  title = {Swing: {{Short-cutting Rings}} for {{Higher Bandwidth Allreduce}}},
  shorttitle = {Swing},
  booktitle = {21st {{USENIX Symposium}} on {{Networked Systems Design}} and {{Implementation}} ({{NSDI}} 24)},
  author = {Sensi, Daniele De and Bonato, Tommaso and Saam, David and Hoefler, Torsten},
  year = 2024,
  pages = {1445--1462},
  publisher = {USENIX Association},
  urldate = {2024-05-30},
  isbn = {978-1-939133-39-7}
}

@inproceedings{shahTACCLGuidingCollective2023,
  address = {Boston, MA, USA},
  title = {{TACCL}: Guiding Collective Algorithm Synthesis Using Communication Sketches},
  shorttitle = {{TACCL}},
  booktitle = {20th {{USENIX Symposium}} on {{Networked Systems Design}} and {{Implementation}} ({{NSDI}} 23)},
  author = {Shah, Aashaka and Chidambaram, Vijay and Cowan, Meghan and Maleki, Saeed and Musuvathi, Madan and Mytkowicz, Todd and Nelson, Jacob and Saarikivi, Olli and Singh, Rachee},
  year = 2023,
  pages = {593--612},
  publisher = {USENIX Association},
  urldate = {2024-06-04},
  isbn = {978-1-939133-33-5}
}

@inproceedings{wangBlinkFastGeneric2020,
  address = {Austin, TX, USA},
  title = {Blink: {{Fast}} and {{Generic Collectives}} for {{Distributed ML}}},
  shorttitle = {Blink},
  booktitle = {Conference on {{Machine Learning}} and {{Systems}} ({{MLSys}} 2020)},
  author = {Wang, Guanhua and Venkataraman, Shivaram and Phanishayee, Amar and Thelin, Jorgen and Devanur, Nikhil and Stoica, Ion},
  year = 2020,
  numpages = {15},
  publisher = {MLSys},
  urldate = {2024-01-11}
}

@inproceedings{xutingliuRethinkingMachineLearning2024,
  title = {Rethinking {{Machine Learning Collective Communication}} as a {{Multi-Commodity Flow Problem}}},
  shorttitle = {{{TE-CCL}}},
  booktitle = {Proceedings of the {{ACM SIGCOMM}} 2024 {{Conference}}},
  author = {{Xuting Liu} and Arzani, Behnaz and Kakarla, Siva Kesava Reddy and Zhao, Liangyu and Liu, Vincent and Castro, Miguel and Kandula, Srikanth and Marshall, Luke},
  year = 2024,
  series = {{{ACM SIGCOMM}} '24},
  pages = {16--37},
  publisher = {Association for Computing Machinery},
  address = {New York, NY, USA},
  urldate = {2024-08-18},
  isbn = {979-8-4007-0614-1}
}

@misc{PortSplittingConfigurations,
  author = {{NVIDIA}},
  key = {NVIDIA},
  title = {Port {{Splitting Configurations}} - {{NVIDIA Docs}}},
  url = {https://docs.nvidia.com/networking/display/connectx8supernic/port-splitting-configurations},
  urldate = {2025-12-01},
  year  = {2025}
}

@misc{BroadcomEthernetNetwork,
  author = {{Broadcom}},
  key = {Broadcom},
  title = {Broadcom {{Ethernet Network Adapter User~Guide}}},
  url = {https://techdocs.broadcom.com/us/en/storage-and-ethernet-connectivity/ethernet-nic-controllers/bcm957xxx/adapters.html},
  urldate = {2025-12-01},
  year  = {2025}
}

@misc{IntelEthernet800,
  author = {{Intel}},
  key = {Intel},
  title = {{{Intel}}® {{Ethernet}} 800 {{Series Product Guide}}},
  url = {https://www.intel.com/content/www/us/en/content-details/709766/intel-ethernet-800-series-product-guide.html},
  urldate = {2025-12-01},
  year  = {2025}
}

@misc{EthernetAdaptersControllers,
  author = {{Marvell}},
  key = {Marvell},
  title = {Ethernet {{Adapters}} and {{Controllers}} - {{FastLinQ Performance NICs}} - 45000 {{FastLinQ CNA}} - {{Marvell}}},
  url = {https://www.marvell.com/products/ethernet-adapters-and-controllers/fastlinq-cna-adapters/45000-fastlinq-cna/documents.html},
  urldate = {2025-12-01},
  year  = {2025}
}

@article{sandersTwotreeAlgorithmsFull2009,
  title = {Two-Tree Algorithms for Full Bandwidth Broadcast, Reduction and Scan},
  shorttitle = {{{DBT}}({{Double Binary Tree}})},
  author = {Sanders, Peter and Speck, Jochen and Träff, Jesper Larsson},
  journal = {Parallel Computing},
  volume = {35},
  number = {12},
  pages = {581--594},
  doi = {10.1016/j.parco.2009.09.001},
  issn = {0167-8191},
  url = {https://doi.org/10.1016/j.parco.2009.09.001},
  urldate = {2025-05-21},
  year  = {2009}
}

@techreport{mitchellPulpLinearProgramming2011,
  title = {Pulp: A Linear Programming Toolkit for Python},
  shorttitle = {Pulp},
  author = {Mitchell, Stuart and O'Sullivan, Michael and Dunning, Iain},
  institution = {Department of Engineering Science, The University of Auckland},
  address = {Auckland, New Zealand},
  year = {2011},
  month = sep,
  url = {https://www.dit.uoi.gr/e-class/modules/document/file.php/216/PAPERS/2011.%20PuLP%20-%20A%20Linear%20Programming%20Toolkit%20for%20Python.pdf},
  urldate = {2024-04-21}
}

@article{hockney1994communication,
  title = {The Communication Challenge for {{MPP}}: {{Intel}} Paragon and Meiko {{CS-2}}},
  shorttitle = {Hockney Model},
  author = {Hockney, Roger W},
  year = {1994},
  journal = {Parallel Computing},
  volume = {20},
  number = {3},
  pages = {389--398},
  publisher = {Elsevier}
}

@inproceedings{wuACTINAAdaptingCircuitSwitching2025,
  address = {New York, NY, USA},
  title = {{{ACTINA}}: {{Adapting Circuit-Switching Techniques}} for {{AI Networking Architectures}}},
  shorttitle = {{{ACTINA}}},
  booktitle = {Proceedings of the {{International Conference}} for {{High Performance Computing}}, {{Networking}}, {{Storage}} and {{Analysis}}},
  author = {Wu, Zhenguo and Klenk, Benjamin and Dennison, Larry and Bergman, Keren},
  year = {2025},
  series = {{{SC}} '25},
  pages = {1211--1222},
  publisher = {Association for Computing Machinery},
  url = {https://dl.acm.org/doi/10.1145/3712285.3759842},
  urldate = {2025-11-16},
  isbn = {979-8-4007-1466-5}
}
\appendix
\clearpage
\section{Execution Model Details}
\label{app:execution-model-details}

\hlmone{This appendix expands the execution contract summarized in \S\ref{subsec: framework}.
The offline scheduler still makes the two optimization decisions in \S\ref{sec:schedule}: the traffic allocation $d_{i,j}$ and the reconfiguration decision $r_{i,j}$.
For execution, the resulting schedule is interpreted as a set of \textit{guarded operations} on each OCS plane, not as a list of absolute-time alarms.
Each operation records its logical step, assigned plane, required configuration, data volume, and dependencies; the two operation types used below are ``path preparation'' on an OCS plane and ``data transfer'' through a prepared path.}

\hlmone{The runtime contract uses four logical events, written as event predicates.
\texttt{path-ready}$(j,\text{cfg}_i)$ means OCS plane $j$ already holds, or has finished preparing, the configuration required by step $i$.
\texttt{plane-free}$(j)$ means no earlier operation on plane $j$ conflicts with the next guarded operation.
\texttt{data-ready}$(i)$ means the previous collective step has produced the data needed by step $i$.
\texttt{transfer-complete}$(i,j)$ means the transfer fragment for step $i$ on plane $j$ has finished and can advance later guards.
In the order of the scheduling properties in \S\ref{sec:schedule}, \texttt{path-ready}$(j,\text{cfg}_i)$ enforces transmission-reconfiguration precedence \hyperref[prop:p1]{(P1)}, \texttt{plane-free}$(j)$ enforces serial use of each OCS plane \hyperref[prop:p2]{(P2)}, and \texttt{data-ready}$(i)$ enforces the logical step barrier \hyperref[prop:p3]{(P3)}; \texttt{transfer-complete}$(i,j)$ is the feedback event that updates the dependency state for later \texttt{plane-free}$(j)$ and \texttt{data-ready}$(i)$ events.
These events can be supplied by existing mechanisms such as RDMA completions, CCL/MPI progress, a root/coordinator process, or controller acknowledgements; the model only relies on their logical meaning, not on a specific implementation.}

\hlmone{The two guarded operations use these predicates differently.
``Path preparation'' is a controller-side action: after the previous scheduled action on OCS plane $j$ has completed, the controller may prepare or reuse configuration $\text{cfg}_i \in \{\mathbf{P}_1,\mathbf{P}_2,\mathbf{P}_3\}$ and then publish \texttt{path-ready}$(j,\text{cfg}_i)$.
Because path preparation sends no collective data, it need not wait for \texttt{data-ready}$(i)$.
``Data transfer'' is a host-side action: it sends $d_{i,j}$ only after \texttt{path-ready}$(j,\text{cfg}_i)$, \texttt{plane-free}$(j)$, and \texttt{data-ready}$(i)$ all hold.
This lets \ours overlap reconfiguration with earlier transfers while preserving the step dependencies of the collective algorithm.}

\suppressfloats[t]

\begin{figure}[H]
  \centering
  \definecolor{swotbluefill}{HTML}{eaf3ff}
  \definecolor{swotblueline}{HTML}{6b8fc2}
  \definecolor{swotgreenfill}{HTML}{eef7ec}
  \definecolor{swotgreenline}{HTML}{76a66f}
  \definecolor{swotshimfill}{HTML}{fff2cc}
  \definecolor{swotshimline}{HTML}{d6b656}
  \definecolor{swotctrlfill}{HTML}{ffe6cc}
  \definecolor{swotctrlline}{HTML}{d79c00}
  \def\guardfield#1{\fcolorbox{swotgreenline}{white}{\scriptsize\strut\textsf{#1}}}
  \resizebox{0.97\textwidth}{!}{%
  \begin{tikzpicture}[
    font=\footnotesize,
    schedbox/.style={draw=swotblueline, fill=swotbluefill, align=center, inner sep=5pt, minimum height=11mm, text width=2.55cm},
    box/.style={draw=swotblueline, fill=swotbluefill, rounded corners, align=center, inner sep=5pt, minimum height=11mm, text width=2.55cm},
    shimbox/.style={draw=swotshimline, fill=swotshimfill, rounded corners, align=center, inner sep=4pt, minimum height=11mm, text width=2.6cm},
    ctrlbox/.style={draw=swotctrlline, fill=swotctrlfill, rounded corners, align=center, inner sep=4pt, minimum height=11mm, text width=2.6cm},
    guard/.style={draw=swotgreenline, fill=swotgreenfill, rounded corners, align=left, inner sep=5pt, minimum height=18mm, text width=4.1cm},
    arrow/.style={-{Latex[length=2mm]}, semithick},
    controllink/.style={-{Latex[length=1.7mm]}, dashed, gray!70, semithick},
    dashedarrow/.style={-{Latex[length=1.8mm]}, dashed, gray!70}
  ]
    \node[schedbox] (sched) at (0,2.0) {\textbf{Offline scheduler}\\$d_{i,j}$, $r_{i,j}$};
    \node[box, text width=3.2cm] (pkg) at (0,0) {\textbf{Global schedule package}\\transfer plan $d_{i,j}$\\config plan $\text{cfg}_i,r_{i,j}$\\dependency contract};
    \node[shimbox] (shim) at (3.7,1.45) {\textbf{Host-side shims}\\NIC workers\\transfer plan};
    \node[ctrlbox] (ctrl) at (3.7,-1.45) {\textbf{Optical controller}\\OCS coordinator\\config plan};
    \node[guard] (xfer) at (8.0,1.7) {{\centering\textbf{Data transfer guard}\par}
      \guardfield{Inputs} \texttt{path-ready}$(j,\text{cfg}_i)$, \texttt{plane-free}$(j)$, \texttt{data-ready}$(i)$\\[1pt]
      \guardfield{Action} issue transfer $d_{i,j}$\\[1pt]
      \guardfield{Output} \texttt{transfer-complete}$(i,j)$\\[1pt]
      \guardfield{Ensures} \hyperref[prop:p1]{(P1)}--\hyperref[prop:p3]{(P3)}};
    \node[guard] (prepare) at (8.0,-1.7) {{\centering\textbf{Path preparation guard}\par}
      \guardfield{Inputs} \texttt{plane-free}$(j)$, prev op\\[1pt]
      \guardfield{Action} prepare $\text{cfg}_i$ on plane $j$\\[1pt]
      \guardfield{Output} \texttt{path-ready}$(j,\text{cfg}_i)$\\[1pt]
      \guardfield{Ensures} \hyperref[prop:p1]{(P1)}, \hyperref[prop:p2]{(P2)}};

    \draw[arrow] (sched.south) -- (pkg.north);
    \draw[arrow] (pkg.north east) -- (shim.west);
    \draw[arrow] (pkg.south east) -- (ctrl.west);
    \draw[controllink] (shim.east) -- (xfer.west);
    \draw[controllink] (ctrl.east) -- (prepare.west);
    \draw[dashedarrow] (prepare.east) to[out=0,in=-35] node[right, align=left, font=\scriptsize] {\texttt{path-ready}$(j,\text{cfg}_i)$\\event signal} (xfer.east);
  \end{tikzpicture}}
  \caption{\textbf{Logical execution contract generated from the offline schedule.}
  The offline scheduler produces one global schedule package, which fans out to host-side shims/NIC workers and the optical controller/OCS coordinator.
  The dashed links from runtime components to guard cards denote which side evaluates each guard, not an additional execution-order constraint.
  The path-preparation guard turns an idle plane and a completed prior action on that plane into \texttt{path-ready}$(j,\text{cfg}_i)$, preserving transmission-reconfiguration precedence \hyperref[prop:p1]{(P1)} and serial plane use \hyperref[prop:p2]{(P2)}.
  The data-transfer guard issues a transfer only when \texttt{path-ready}$(j,\text{cfg}_i)$, \texttt{plane-free}$(j)$, and \texttt{data-ready}$(i)$ hold, enforcing properties \hyperref[prop:p1]{(P1)}--\hyperref[prop:p3]{(P3)} and emitting \texttt{transfer-complete}$(i,j)$ for later guards.}
  \Description{A logical execution diagram. The offline scheduler sits above and feeds a global schedule package with a transfer plan, configuration plan, and dependency contract. The package fans out with straight schedule-distribution arrows to host-side shims or NIC workers and to the optical controller or OCS coordinator. Dashed control links connect the host-side shims to the data-transfer guard and the optical controller to the path-preparation guard; these links indicate guard evaluation responsibility, not execution order. Path preparation is guarded by plane-free and completion of the prior action on that plane and produces path-ready for the plane and configuration. Data transfer is guarded by path-ready, plane-free, and data-ready and produces transfer-complete. A dashed event-signal arrow carries path-ready from the path-preparation guard to the data-transfer guard.}
  \label{fig:execution-contract}
  \vspace{-0.15in}
\end{figure}

\hlmone{Figure~\ref{fig:execution-contract} summarizes the runtime views and guards; the complete logical contract is as follows.
At initialization, \texttt{data-ready}$(1)$ and \texttt{plane-free}$(j)$ are satisfied, and any initial OCS configuration needed by the first assigned transfer is treated as an already satisfied \texttt{path-ready} event.}

\begin{enumerate}[leftmargin=*, itemsep=0pt, topsep=2pt, parsep=0pt]
  \item \hlmone{\textbf{Schedule generation.}
  The offline scheduler computes the traffic allocation $d_{i,j}$ and reconfiguration decision $r_{i,j}$, together with an ordering that satisfies properties \hyperref[prop:p1]{(P1)}--\hyperref[prop:p3]{(P3)} in \S\ref{sec:schedule}.}
  \item \hlmone{\textbf{Plan distribution.}
  The resulting schedule package is split into a transfer view, a configuration view, and a dependency contract.
  These runtime views and the dependency state derived from them are sufficient to determine which operation can be released next on each plane.}
  \item \hlmone{\textbf{Path readiness.}
  If a future transfer requires a different configuration, the plane can prepare that path after the previous scheduled action on the same OCS plane completes.
  Once the path is prepared, the model emits \texttt{path-ready}$(j,\text{cfg}_i)$; if the current configuration already matches the required $\text{cfg}_i$, \texttt{path-ready}$(j,\text{cfg}_i)$ is already satisfied.}
  \item \hlmone{\textbf{Completion aggregation.}
  The data-transfer guard emits \texttt{transfer-complete}$(i,j)$ when a released fragment finishes, and these completions update the dependency state.
  When all active fragments of step $i-1$ complete, the model emits \texttt{data-ready}$(i)$, implementing the logical step barrier without synchronized clock firing.}
  \item \hlmone{\textbf{Guarded issue.}
  A transfer on plane $j$ for step $i$ is issued only when $d_{i,j}>0$, \texttt{path-ready}$(j,\text{cfg}_i)$, \texttt{plane-free}$(j)$, and \texttt{data-ready}$(i)$ are all satisfied.}
  \item \hlmone{\textbf{Feedback.}
  If a transfer or path-preparation operation takes longer than planned, the corresponding event simply arrives later; dependent guards remain closed until it arrives.
  This shifts later releases and reduces realized overlap, but it does not invalidate the plan or violate properties \hyperref[prop:p1]{(P1)}--\hyperref[prop:p3]{(P3)}.}
\end{enumerate}

\hlmone{Figure~\ref{fig:execution-example} instantiates this contract for the overlap schedule in Figure~\ref{fig: moti_example}(b).
It preserves the same transfer and reconfiguration blocks, but redraws their execution as a guarded event trace: logical events satisfy guard inputs, and satisfied guards release the next operation without requiring synchronized wall-clock firing.}

\begin{figure}[H]
  \centering
  \definecolor{swotreconffill}{HTML}{fff2cc}
  \definecolor{swotreconfline}{HTML}{d6b657}
  \definecolor{swotreconftext}{HTML}{8a7300}
  \definecolor{swotguardfill}{HTML}{eef7ec}
  \definecolor{swotguardline}{HTML}{76a66f}
  \definecolor{swoteventfill}{HTML}{eaf3ff}
  \definecolor{swoteventline}{HTML}{6b8fc2}
  \definecolor{swotretainfill}{HTML}{f5f5f5}
  \resizebox{0.98\textwidth}{!}{%
  \begin{tikzpicture}[
    font=\footnotesize,
    txstep/.style={draw=black, fill=white, align=center, minimum height=8.2mm, inner sep=3pt},
    reconf/.style={draw=swotreconfline, fill=swotreconffill, text=swotreconftext, align=center, minimum height=8.2mm, inner sep=3pt},
    retain/.style={draw=gray!65, dashed, fill=swotretainfill, align=center, minimum height=8.2mm, inner sep=3pt},
    event/.style={draw=swoteventline, fill=swoteventfill, rounded corners, align=center, inner sep=2.5pt, text width=2.0cm, font=\scriptsize},
    guard/.style={draw=swotguardline, fill=swotguardfill, rounded corners, align=center, inner sep=3pt, text width=2.35cm, font=\scriptsize},
    smallguard/.style={draw=swotguardline, fill=swotguardfill, rounded corners, align=center, inner sep=2.5pt, text width=1.55cm, font=\scriptsize},
    lgtx/.style={draw=black, fill=white, minimum width=3.8mm, minimum height=2.6mm, inner sep=0pt},
    lgreconf/.style={draw=swotreconfline, fill=swotreconffill, minimum width=3.8mm, minimum height=2.6mm, inner sep=0pt},
    lgretain/.style={draw=gray!65, dashed, fill=swotretainfill, minimum width=3.8mm, minimum height=2.6mm, inner sep=0pt},
    lgevent/.style={draw=swoteventline, fill=swoteventfill, rounded corners, minimum width=3.8mm, minimum height=2.6mm, inner sep=0pt},
    lgguard/.style={draw=swotguardline, fill=swotguardfill, rounded corners, minimum width=3.8mm, minimum height=2.6mm, inner sep=0pt},
    legendtext/.style={font=\scriptsize, anchor=west},
    arrow/.style={-{Latex[length=1.8mm]}, semithick},
    dashedarrow/.style={-{Latex[length=1.5mm]}, dashed, gray!70}
  ]
    \node[lgtx] at (0.15,3.05) {};
    \node[legendtext] at (0.42,3.05) {transfer};
    \node[lgreconf] at (1.85,3.05) {};
    \node[legendtext] at (2.12,3.05) {path prep};
    \node[lgretain] at (3.75,3.05) {};
    \node[legendtext] at (4.02,3.05) {retain};
    \node[lgevent] at (5.35,3.05) {};
    \node[legendtext] at (5.62,3.05) {event};
    \node[lgguard] at (6.85,3.05) {};
    \node[legendtext] at (7.12,3.05) {guard};
    \draw[dashedarrow] (8.25,3.05) -- (8.85,3.05);
    \node[legendtext] at (8.95,3.05) {event signal};
    \draw[arrow] (10.65,3.05) -- (11.25,3.05);
    \node[legendtext] at (11.35,3.05) {release};

    \node[anchor=east] at (-0.25,0.45) {$OCS_1$};
    \node[anchor=east] at (-0.25,-0.65) {$OCS_2$};

    \node[txstep, anchor=west, minimum width=3.0cm] (o1s1) at (0,0.45) {Step 1\\15MB, $\mathbf{P}_1$};
    \node[reconf, anchor=west, minimum width=2.0cm] (o1r13) at (3,0.45) {$\mathbf{P}_1{\rightarrow}\mathbf{P}_3$};
    \node[txstep, anchor=west, minimum width=1.0cm] (o1s3) at (5,0.45) {Step 3\\5MB};
    \node[txstep, anchor=west, minimum width=1.0cm] (o1s4) at (6,0.45) {Step 4\\5MB};
    \node[reconf, anchor=west, minimum width=2.0cm] (o1r31) at (7,0.45) {$\mathbf{P}_3{\rightarrow}\mathbf{P}_1$};
    \node[txstep, anchor=west, minimum width=3.0cm] (o1s6) at (9,0.45) {Step 6\\15MB};

    \node[txstep, anchor=west, minimum width=1.0cm] (o2s1) at (0,-0.65) {Step 1\\5MB};
    \node[reconf, anchor=west, minimum width=2.0cm] (o2r12) at (1,-0.65) {$\mathbf{P}_1{\rightarrow}\mathbf{P}_2$};
    \node[txstep, anchor=west, minimum width=2.0cm] (o2s2) at (3,-0.65) {Step 2\\10MB};
    \node[retain, anchor=west, minimum width=2.0cm] (o2idle) at (5,-0.65) {retain $\mathbf{P}_2$\\skip $\mathbf{P}_3$};
    \node[txstep, anchor=west, minimum width=2.0cm] (o2s5) at (7,-0.65) {Step 5\\10MB};
    \node[reconf, anchor=west, minimum width=2.0cm] (o2r21) at (9,-0.65) {$\mathbf{P}_2{\rightarrow}\mathbf{P}_1$};
    \node[txstep, anchor=west, minimum width=1.0cm] (o2s6) at (11,-0.65) {Step 6\\5MB};

    \node[event] (e2free) at (1.25,2.0) {\texttt{plane-free}$(2)$\\local Step~1 done};
    \node[guard, text width=2.55cm] (g2) at (3.85,2.05) {$G_2$ transfer guard\\\texttt{path-ready}$(2,\mathbf{P}_2)$\\\texttt{plane-free}$(2)$\\\texttt{data-ready}$(2)$};
    \node[event] (ep3) at (6.25,2.0) {\texttt{path-ready}$(1,\mathbf{P}_3)$\\bypass $\mathbf{P}_2$};
    \node[guard, text width=2.15cm] (g5) at (7.15,-2.25) {$G_5$ transfer guard\\\texttt{data-ready}$(5)$\\retained $\mathbf{P}_2$\\\texttt{plane-free}$(2)$};
    \node[event, text width=1.65cm] (edata6) at (9.2,-2.25) {\texttt{data-ready}$(6)$\\Step~5 done};
    \node[smallguard] (g61) at (9.8,1.78) {$G_{6,1}$\\\texttt{path-ready}\\\texttt{data-ready}};
    \node[smallguard] (g62) at (11.25,-2.25) {$G_{6,2}$\\\texttt{path-ready}\\\texttt{data-ready}};

    \draw[dashedarrow] (o2s1.north) -- (e2free.south);
    \draw[arrow] (e2free.south) to[out=-90,in=95] (o2r12.north);
    \draw[dashedarrow] (o1s1.north east) to[out=75,in=205] (g2.west);
    \draw[dashedarrow] (o2r12.north east) to[out=70,in=250] (g2.south);
    \draw[arrow] (g2.south) to[out=-90,in=95] (o2s2.north);

    \draw[dashedarrow] (o1r13.north east) -- (ep3.south);
    \draw[dashedarrow] (o1s4.south east) to[out=-70,in=155] (g5.west);
    \draw[dashedarrow] (o2idle.south east) to[out=-80,in=115] (g5.north);
    \draw[arrow] (g5.north) to[out=95,in=-95] (o2s5.south);

    \draw[dashedarrow] (o2s5.south east) to[out=-70,in=155] (edata6.west);
    \draw[dashedarrow] (o1r31.south east) to[out=-70,in=-105] (g61.south);
    \draw[dashedarrow] (edata6.north) to[out=90,in=-105] (g61.south west);
    \draw[arrow] (g61.south) to[out=-90,in=110] (o1s6.north west);
    \draw[dashedarrow] (o2r21.south east) to[out=-70,in=120] (g62.north);
    \draw[dashedarrow] (edata6.east) -- (g62.west);
    \draw[arrow] (g62.north) to[out=90,in=-95] (o2s6.south);
  \end{tikzpicture}}
  \caption{\textbf{Guarded event trace for the overlap schedule in Figure~\ref{fig: moti_example}(b).}
  The blocks are the same transfers, path preparations, and retained configurations as in the motivating example, but no block is assumed to start from a synchronized clock alarm.
  Blue event tokens are emitted by local completion or path readiness; dashed arrows show event signals that satisfy guard inputs or persist as readiness state; solid arrows show a satisfied guard releasing an operation.
  For Step~2, $OCS_2$ prepares $\mathbf{P}_2$ after its local Step~1 fragment completes, yet the transfer waits until \texttt{path-ready}$(2,\mathbf{P}_2)$, \texttt{plane-free}$(2)$, and \texttt{data-ready}$(2)$ all hold.
  The same rule releases Step~5 on retained $\mathbf{P}_2$ and releases the two Step~6 fragments through independent per-plane guards.}
  \Description{A guarded event trace for the two-lane OCS timeline from the motivating example. A legend maps white blocks to data transfers, yellow blocks to path preparation, gray dashed blocks to retained configurations, blue rounded boxes to logical events, green rounded boxes to guards, dashed arrows to event signals or readiness state, and solid arrows to released operations. OCS1 executes Step 1 with P1, prepares P3 while bypassing P2, executes Step 3 and Step 4, prepares P1, and executes a Step 6 fragment. OCS2 executes a smaller Step 1 fragment, prepares P2, executes Step 2, retains P2 while skipping P3, executes Step 5, prepares P1, and executes a Step 6 fragment. The Step 2 guard waits for path-ready, plane-free, and data-ready. The Step 5 guard waits for data-ready, retained P2, and plane-free. Two Step 6 guards release the two fragments independently after data-ready and per-plane path readiness.}
  \label{fig:execution-example}
\end{figure}

\hlmone{\textbf{Scope.}
This appendix gives a schedule-level logical execution contract, not a production runtime specification.
The contract assumes reliable readiness/completion notifications, a stable out-of-band control path, and a conservative \texttt{path-ready}$(j,\text{cfg}_i)$ event that may include switch actuation, settling, controller acknowledgement, and hardware-specific link-readiness delay.
Longer-than-planned sends or delayed readiness events push affected guards later, so the realized CCT may increase; however, event-triggered release prevents premature sends or unsafe reconfiguration, so execution degrades in performance rather than failing structurally.
The evaluation studies schedule-level benefits under parameterized bandwidth, communication latency, and conservative path-readiness latency.
Concrete RDMA/CCL/MPI integrations, root/coordinator protocols, production overheads, online fallback, and dynamic replanning remain outside the scope of this paper.}



\end{document}